\documentclass[journal]{IEEEtran}
\usepackage[final]{changes}

\IEEEoverridecommandlockouts

\usepackage{cite}

\usepackage[english]{babel}

\ifCLASSINFOpdf
   \usepackage[pdftex]{graphicx}
\else
   \usepackage[pdftex]{graphicx}
   \DeclareGraphicsExtensions{.eps}
\fi

\usepackage{grffile}
\usepackage{epstopdf}

\usepackage[tight,footnotesize]{subfigure}

\usepackage{fixltx2e}

\usepackage[cmex10]{amsmath}
\usepackage{color}
\usepackage{array}

\begin{document}
%-----------------------------------------------------------------------------------

\title{Random Access in DVB-RCS2: Design and Dynamic Control for Congestion Avoidance}

\author{\IEEEauthorblockN{Alessio Meloni,~\IEEEmembership{Student~Member,~IEEE},\thanks{A. Meloni gratefully acknowledges Sardinia Regional Government for the financial support of his PhD scholarship (P.O.R. Sardegna F.S.E. 2007-2013 - Axis IV Human Resources, Objective l.3, Line of Activity l.3.1.).} Maurizio Murroni\thanks{\copyright 2014 IEEE. The IEEE copyright notice applies. DOI: 10.1109/TBC.2013.2293920},~\IEEEmembership{Senior Member,~IEEE} }

\thanks{The authors are with the Department of Electrical and Electronic Engineering (DIEE), University of Cagliari, Piazza D'Armi, 09123 Cagliari, Italy (e-mail: alessio.meloni@diee.unica.it;\ murroni@diee.unica.it .}
}

\maketitle

\begin{abstract}
In the current DVB generation, satellite terminals are expected to be interactive and capable of transmission in the return channel with satisfying quality. Considering the bursty nature of their traffic and the long propagation delay, the use of a Random Access technique is a viable solution for such a Medium Access Control (MAC) scenario. In this paper Random Access communication design in DVB-RCS2 is considered with particular regard to the recently introduced Contention Resolution Diversity Slotted Aloha (CRDSA) technique. The paper presents a model for design and tackles some issues on performance evaluation of the system by giving intuitive and effective tools. Moreover, dynamic control procedures able to avoid congestion at the gateway are introduced. Results show the advantages brought by CRDSA to DVB-RCS2 with regard to the previous state of the art.
\end{abstract}

\begin{IEEEkeywords}
Retransmission, Successive Interference Cancellation, Slotted Aloha, Random Access, Congestion Avoidance, Dynamic Control, Satellite Broadcasting, Interactive Terminals
\end{IEEEkeywords}

\IEEEpeerreviewmaketitle

\section{Introduction} \label{Intro}

Current DVB standards \added{such as Digital Video Broadcasting over Satellite} are increasingly going towards the need of high-speed bidirectional networks for consumer interactivity \cite{interact}. Nevertheless in consumer type of interactive satellite terminals, users generate a large amount of low duty cycle and bursty traffic with frequent periods of inactivity in the return link. Under these operating conditions, the traditionally used Demand Assignment Multiple Access (DAMA) satellite protocol does not perform optimally, since the response time for the transmission of short bursts can be too long. For this reason, in the recently approved specification for the next generation of Interactive Satellite Systems (DVB-RCS2) \cite{DVB}, the possibility of sending logon, control and even user traffic using Random Access (RA) in timeslots specified by the Network Control Center (NCC) is provided to Return Channel Satellite Terminals (RCST). In particular, two methods are considered for RA: the first one is Slotted Aloha (SA) \cite{roberts} \cite{abramson}, the second is called Contention Resolution Diversity Slotted Aloha (CRDSA) \cite{CRDSA1}.
 
SA represents a well established RA technique for satellite networks in which users send their bursts within slots in a distributed manner, i.e. without any central entity coordinating transmission. This allows to reach an average throughput around $0.36\ [packets/slot]$ despite the possibility of collision among bursts from different users. Nevertheless, in practice SA works with very moderate channel load to ensure acceptable delay and loss probability ($PLR=10^{-3}$ for $10^{-3} [packets/slot]$). 

This gap has been recently filled with the introduction of a new technique named CRDSA that is able to boost the throughput even up to values close to $1  [packets/slot]$. This technique is based on the transmission of a chosen number of replicas for each burst payload as in Diversity Slotted Aloha (DSA) \cite{DSA}. However, differently from DSA, in CRDSA each burst copy has a pointer to the location of the other replicas of the same burst payload so that when the frame arrives at the receiver a Successive Interference Cancellation (SIC) process is accomplished. The SIC process consists in removing the content of already decoded bursts from the slots in which collision occurred. Doing so, it is possible to restore the content of bursts that had all their replicas colliding thus boosting the throughput. Nevertheless the use of interference cancellation can benefit from power unbalance among different sources \cite{interf1} \cite{interf2}.

\added{Considering these recent findings on Aloha-based RA and the fact that in \cite{DVB}}\deleted{The specification for the lower layers of DVB-RCS2 [2] states that the applications using the interactive network may rely on network internal} contention control mechanisms to avoid \deleted{sustained excessive packet loss resulting from simultaneous destructive transmissions (i.e. }congestion \replaced{are}{). For this reason, the definition of such a mechanism is} claimed to be out of scope for the document\replaced{, in}{. In} this paper\deleted{,} a complete model for the design of \deleted{such} a\added{ congestion-free} system\deleted{ and simple yet effective dynamic retransmission policies aiming at avoiding congestion at the gateway are}\added{using CRDSA is} presented. To do so, we rely on the analysis we carried out in \cite{stab1}, in which the\added{ need for such a} model was firstly introduced\deleted{ and throughput results for CRDSA with geometrically distributed retransmissions were analyzed}. 

\deleted{The presented model is based on the definition of \textit{Equilibrium Contour} [10] and \textit{Channel Load Line}. The Equilibrium Contour represents a set of points for different load values, for which the ongoing communication is in equilibrium in the sense that the number of newly transmitted packets is equal to the number of packets successfully sent so that the total number of packets "queuing for retransmission" do not change. The Channel Load Line completes this analysis telling which of these equilibrium points are actually of equilibrium for a given scenario.}

\added{The main contributions of this paper are: the presentation of a stability model and of some crucial evaluation parameters (packet delay and FET) developed in \cite{stab2} for SA and here adapted to CRDSA; a thorough analysis of the design aspects to take into account in such a system; the adaptation of the dynamic control limit policies developed for SA and the analysis of their convenience when compared to static CRDSA design; the extension of all these results to the case of infinite population that is usually not considered even though in some scenarios is the most appropriate population representation.} 

\deleted{After revising the model presented in [10], tools for the computation of some important parameters such as throughput, packet delay and the First Exit Time are given. Furthermore, design guidelines able to ensure that the communication is taking place in a desirable and stable manner are discussed. Finally, dynamic procedures of the control limit type [11] are introduced as a mean to avoid congestion in DVB-RCS2 when using RA and in particular CRDSA as transmission mode. The paper concludes with a thorough comparison between the use of dynamic control policies in SA and CRDSA and with the extension of the presented model to the case of infinite population, that represents a typical DVB-RCS2 scenario in which a big number of terminals are part of the interactive network and can manage multiple packet transmissions.} 

\added{The remainder of this paper is organized as follows. In Section II a brief overview of the CRDSA scheme is given and the congestion problem that arises when introducing retransmissions in such a scheme is stated. Section III presents the stability model used for analysis and design in the subsequent sections and defines the concept of stability. Section IV regards the definition of a Markov chain for packet delay computation and its validation. In Section V the definition of First Exit Time is given and the formulas for its calculation are presented; moreover the adaptation of the model for computation reduction and for the extension to the case of infinite population are discussed. Section VI exploits the model and the evaluation parameters introduced in the previous sections by giving a general overview of how they are used in the design phase. Section VII compares the advantages and disadvantages of various Aloha-based RA techniques and packet degree choices. Finally Section VIII introduces some dynamic control policies already used in \cite{dyn_stab} for SA while Section IX and X discuss their application respectively to the case of finite and infinite population. Section XI concludes the paper.}

\section{CRDSA System Overview}\label{systOV}

Consider a multi-access channel populated by a total number of users $M$. Users are synchronized so that the channel is divided into slots and $N_s$ consecutive slots are grouped in a frame. The probability that at the beginning of a frame an idle user will send a new packet is $p_0$. When a frame starts, users willing to transmit place $d$ copies of the same packet over the $N_s$ slots of that frame. The number of copies $d$ can be either the same for each packet or not. The first case is known as Constant Replication CRDSA (CR-CRDSA); the second case is known as Variable Replication CRDSA (VR-CRDSA). In this second case the number of copies is randomly generated using a pre-determined probability mass function \cite{Liva}.

Packet copies are nothing else that redundant replicas except for the fact that each one contains a pointer to the location of the others. These pointers are used in order to attempt restoring collided packets at the receiver by means of SIC, i.e. by removing the interfering content of already decoded packets thanks to their location's knowledge. Nevertheless, the physical structure of the burst copies used is equal to the one used for dedicated access with respect to the burst construction, the waveform, the timeslot structure and the burst reception.

\begin{figure}[tb!]
\centering
\includegraphics [width=9 cm] {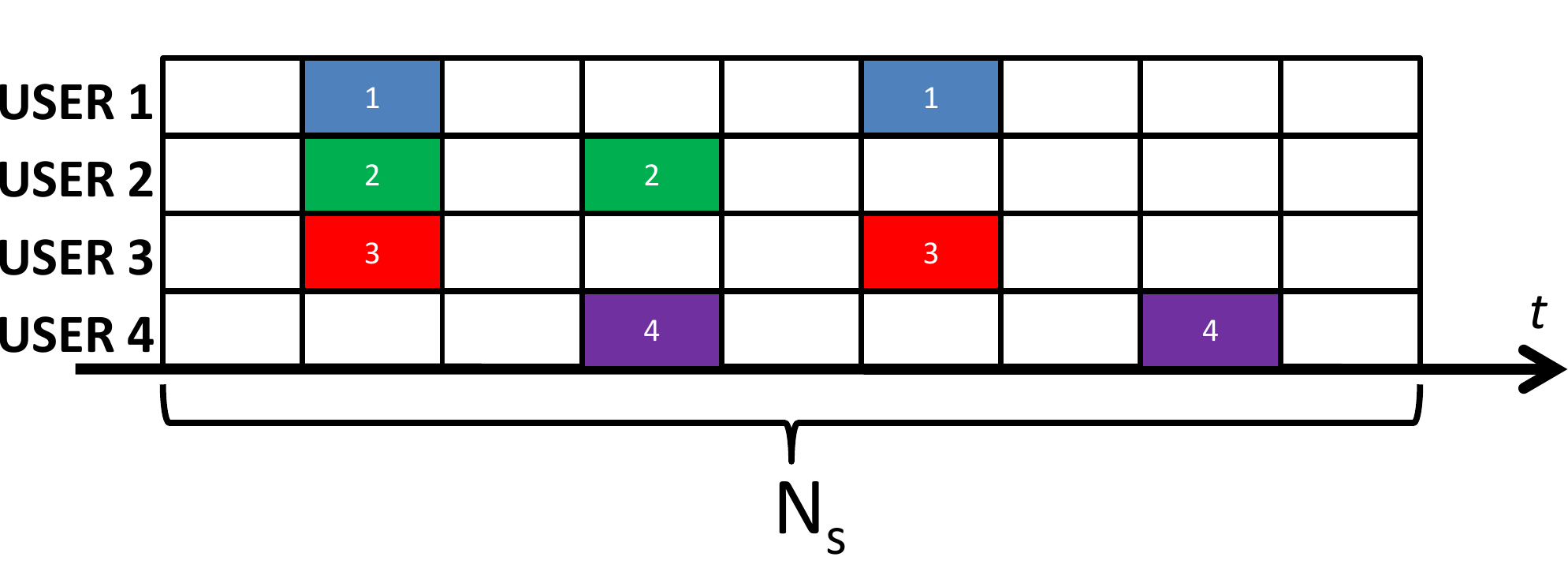}
\caption{Example of frame at the receiver for CRDSA with 2 copies per packet; plain slots indicate that a transmission occurred for that user in that slot.}
\label{Fig1}
\end{figure}

Throughout the paper perfect SIC and ideal channel are assumed, which means that the receiver is always able to perfectly apply SIC without errors and no external sources of disturbance such as noise that might require SNR estimation \cite{SNRest} are present. Therefore, the only cause of unsuccessful decoding of packets is interference among them{\color{black} \footnote{\added{For the case in which non-ideal channel estimation is considered, the reader can refer to \cite{CRDSA1} where it has been demonstrated that there is no appreciable impact on the CRDSA performance for moderate SNR. Moreover, as suggested by one of the reviewers that we kindly thank, we specify here that also situations such as outage that affect the number of backlogged users are not be a problem as long as the channel is nominally stable. In fact, if there is only one point of stability, the channel will rapidly set back to the operational point after the outage period. This sentence is anticipated here but will be clearer in the subsequent sections.}}}. Given these conditions, consider Figure~\ref{Fig1} representing a practical example of SIC. Each slot can be in one of three states:

\begin{itemize}
\item{no packet's copies have been placed in a given slot, thus the slot is empty;}
\item{only 1 packet's copy has been placed in a given slot, thus the packet is correctly decoded;}
\item{more than 1 packet's copy has been placed in a given slot, thus resulting in collision.}
\end{itemize}

In DSA, if all copies \replaced{of}{for} a given packet collided the packet is surely lost. In CRDSA, if at least one copy of a certain packet has been correctly received (see User 4), the contribution of the other copies of the same packet can be removed from the other slots thanks to the knowledge of their location provided by the pointers contained in the decoded packet. This process might allow to restore the content of packets that had all their copies colliding (see User 2) and iteratively other packets may be correctly decoded up to a point in which no more packets can be restored (see User 1 and 3) or until the maximum number of iterations $I_{max}$ for the SIC process is reached, if a limit has been fixed.

If a packet has not been successfully decoded at the end of the SIC process and a retransmission request is sent by the NCC, a feedback is created between the transmitter and the receiver side. This \added{feedback} could cause the channel to get to a state of congestion in \replaced{which}{case} the load is too big \added{because of too many packets to retransmit} \replaced{and causes}{since} excessive packet loss due to simultaneous destructive transmissions\deleted{ of the terminals in the return channel}. For this reason, a careful design of the \replaced{system}{communication} based on a stability model \deleted{and the possible application of control policies able to avoid congestion are}\added{is} of great interest for practical DVB-RCS applications.

\section{Stability Model}

Consider the RA communication system presented in Section~\ref{systOV}. Each user can be in one of two states: Thinking (T) or Backlogged (B). Users in T state generate a new packet for transmission with probability $p_0$ over a frame interval; if so, no other packets are generated until successful transmission for that packet has been acknowledged. Users in B state have unsuccessfully transmitted their packet and keep attempting to retransmit it with probability $p_r$ over a frame interval. In the followings, we assume that users are acknowledged about the success of their transmission at the end of the frame in which the packet has been transmitted (i.e. immediate feedback). Nevertheless this constraint will be relaxed in the last sections.

\subsection{\added{Equilibrium Contour}}

Defining
\begin{itemize}
\item{$N_B^f$ : backlogged users at the end of frame $f$}
\vspace{0.2cm}
\item{$G_B^f=\frac{N_B^{(f-1)}  p_r}{N_s}$ : expected channel load{\color{black}\footnote {\added{In this paper, with channel load we refer to the logical channel load that is the number of packets sent normalized over the frame size, regardless of the number of copies sent per packet.}}} of frame $f$ due to users in B state}
\vspace{0.2cm}
\item{$G_T^f$ : expected channel load of frame $f$ due to users in T state}
\vspace{0.2cm}
\item{$G_{IN}^f=G_T^f+G_B^f$ : expected total channel load of frame $f$}
\vspace{0.2cm}
\item{$PLR^f(G_{IN}^f,N_s,d,I_{max})$ : expected packet loss ratio of frame $f$}
\vspace{0.2cm}
\item{$G_{OUT}^f=G_{IN}^f ( 1 - PLR(G_{IN}^f,N_s,d,I_{max}) )$ : part of load successfully transmitted in frame $f$, i.e. throughput.}
\end{itemize}
\vspace{0.2cm}

The equilibrium contour can be written as

\begin{equation}\label{gt}
 G_T=G_{OUT}=G_{IN} ( 1 - PLR(G_{IN},N_s,d,I_{max}) )
\end{equation}

that represents the \textit{locus of points for which at any time the expected channel load due to users in T state is equal to the expected throughput $G_{OUT}$}. Notice that the frame number $f$ has been omitted, since in equilibrium state this condition is expected to hold for any frame. 

From the definition above we can also gather that the expected number of backlogged users remains the same frame after frame. Therefore

\begin{equation}\label{nb2}
N_B=\frac{G_{IN} PLR(G_{IN},N_s,d,I_{max}) N_s}{p_r}
\end{equation}

Equations \eqref{gt} and \eqref{nb2} completely describe the \textit{equilibrium contour} on the ($G_T$,$N_B$) plane for different values of $G_{IN}${\color{blue}\footnote{Concerning the values used for the Packet Loss Ratio, it is known from the literature \cite{CRDSA1} that the relation between $PLR(G_{IN})$ and $G_{IN}$ can not be \replaced{tightly described}{easily modelled} in an analytical manner. For this reason PLR values used in this work are taken from simulations. \added{However, once simulations for the open loop case are accomplished, they can be used in the stability model to see how the performance changes when modifying some crucial parameters such as $p_r$, $p_0$ and $M$ without the need of running brand new simulations. Moreover simulated values can be substituted with tight analytical equations in case any will be found in the future.}}}.

\subsection{\added{Channel load line}}

Notice that the equilibrium contour itself is an infinite set of equilibrium points. In fact, we have not considered $M$ and $p_0$ so far. This two essential parameters constitute the channel load line and provide the knowledge of the actual stability points for a certain scenario. 

Consider $M$ and $p_0$ to be constant (i.e. stationary input). Given a certain number of backlogged packets, the channel load line expresses the expected value of channel load due to users in T state. For the finite population case, the \textit{channel load line} can be defined as
\begin{equation}\label{LL1}
G_T=\frac{M-N_B}{N_s}p_0
\end{equation}
while for $M\rightarrow\infty$ the channel input can be described as a Poisson process with expected value $\lambda$ [thinking users]  \cite{abramson} so that 
\begin{equation}\label{LL2}
G_T=\frac{\lambda}{N_s}
\end{equation}
 for any $N_B$ (i.e. the expected channel input is constant and independent of the number of backlogged packets).

%%% %%%%%%%
\begin{figure}[tbh!]
\subfigure [Stable channel] {\label{stable} \includegraphics [ scale = 0.2 ]{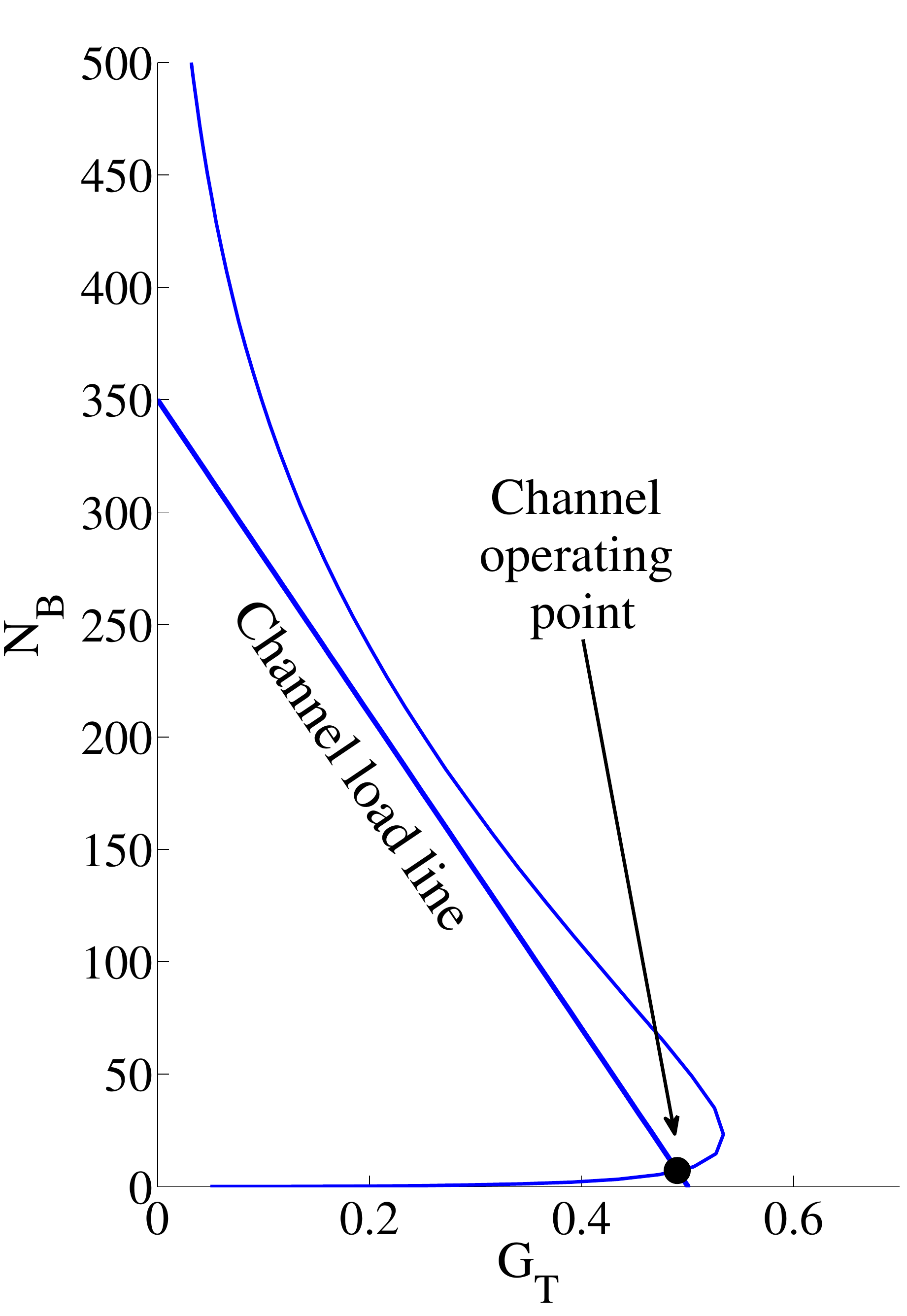}
\label{st}
} \qquad
\subfigure [Unstable channel (finite M)] {\label{unstableFin} \includegraphics [ scale = 0.2 ]{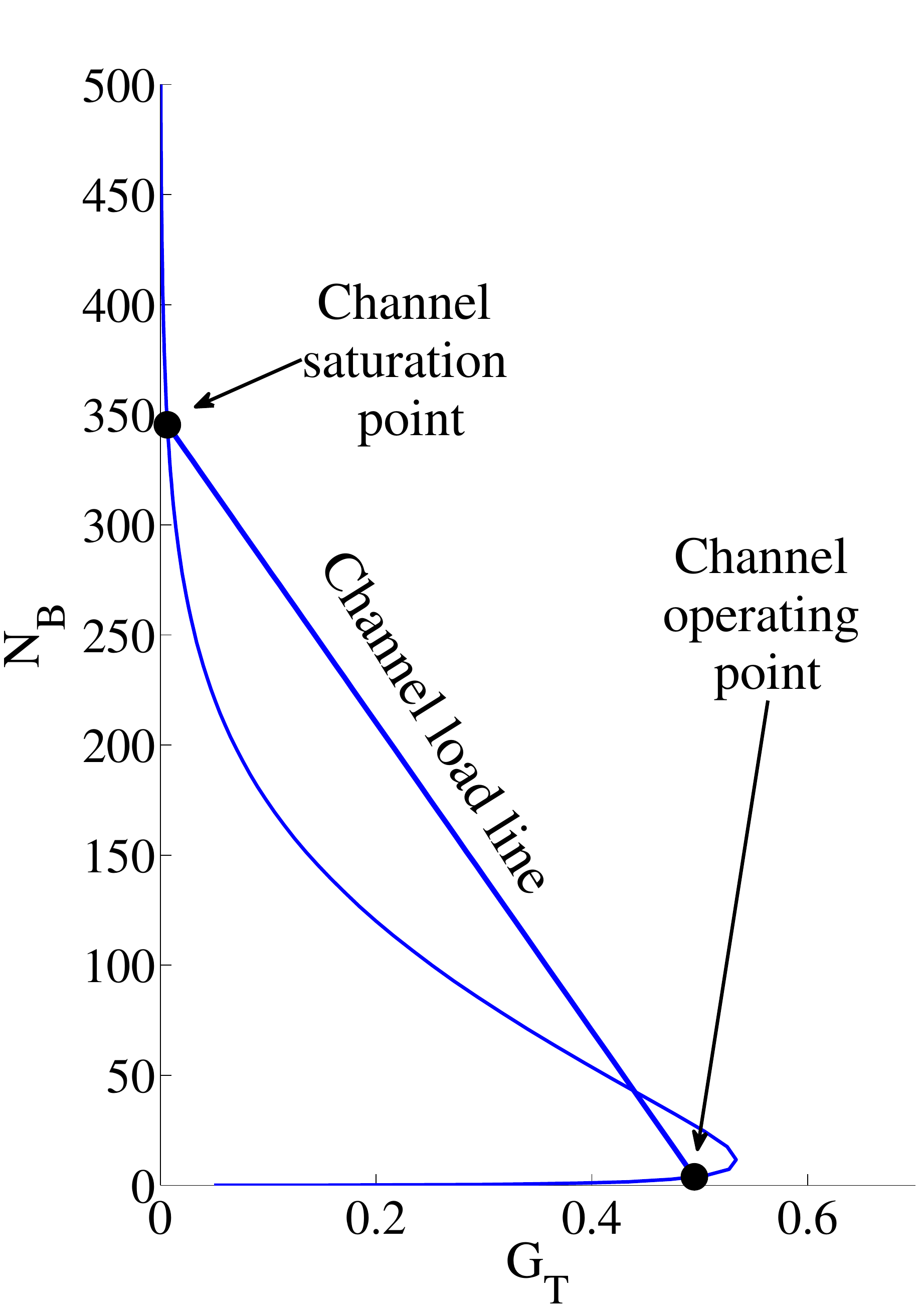}
\label{unstFin}
} \qquad

%\begin{center}
\begin{tabular}{l l}
  $p_0=0.143$ & $p_0=0.143$ \\
\vspace{0.2cm}
  $p_r=0.5$ & $p_r=1$\\
\vspace{0.2cm}
  $M=350$ & $M=350$\\
\vspace{0.2cm}
  $(G_T^G,N_B^G)=(0.49,8.2) \hspace{0.7cm}$ & $(G_T^{S1},N_B^{S1})=(0.495,3.9)$\\
\vspace{0.2cm}
  & $(G_T^U,N_B^U)=(0.44,42.7)$\\
\vspace{0.2cm}
  & $(G_T^{S2},N_B^{S2})=(6\cdot10^{-3},346)$\\
\hline \\

\end{tabular}
%\end{center}

%\ContinuedFloat
\subfigure [Unstable channel (infinite M)] {\label{unstableInf} \includegraphics [ scale = 0.2 ]{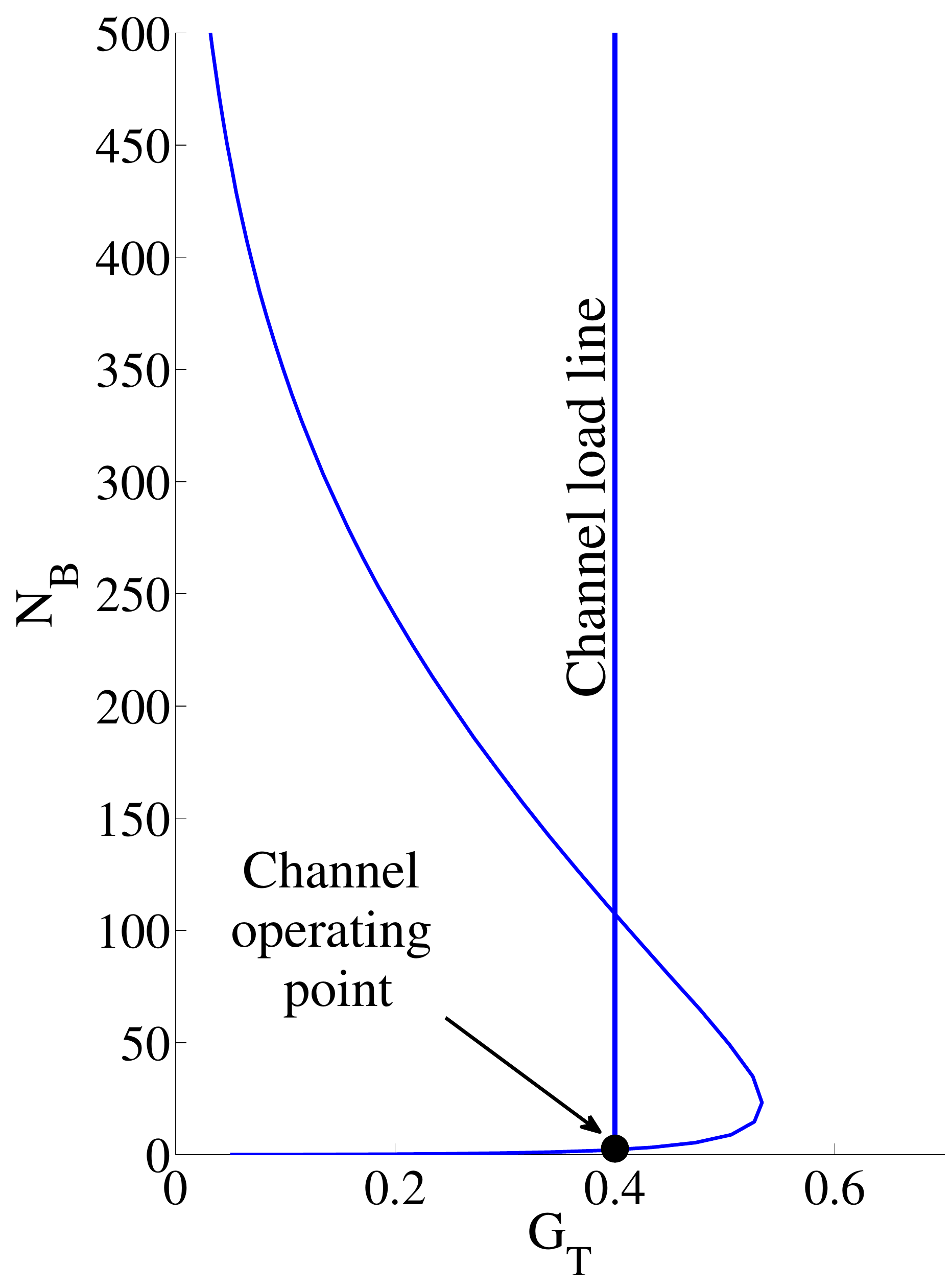}
\label{unstInf}
} \qquad
\subfigure [Overloaded channel] {\label{overL} \includegraphics [ scale = 0.2 ]{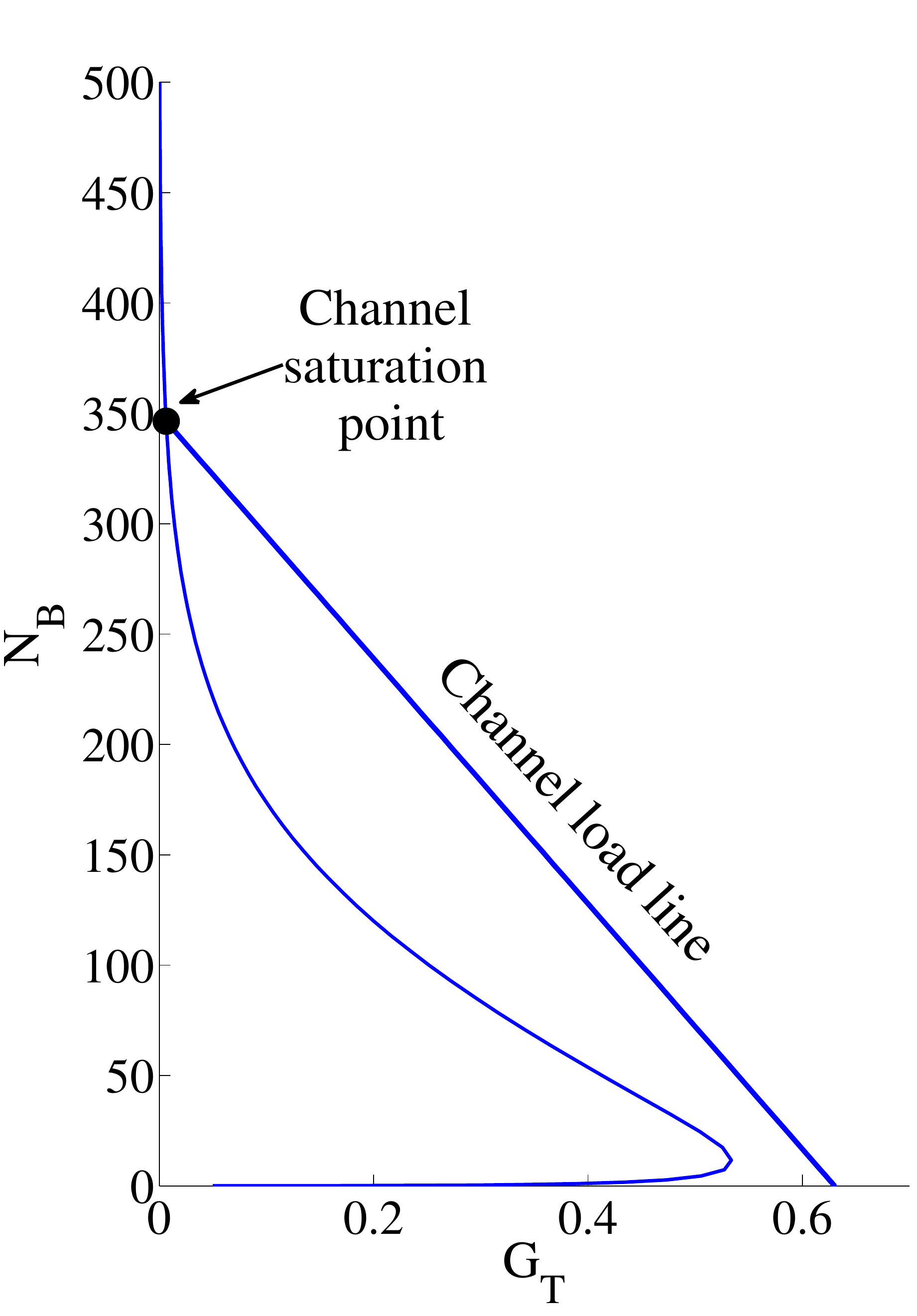}
\label{over}
} \qquad

%\begin{center}
\begin{tabular}{l l}

  $\lambda=40$ & $p_0=0.18$ \\
\vspace{0.2cm}
  $p_r=0.5$ & $p_r=1$\\
\vspace{0.2cm}
  $M\rightarrow\infty$ & $M=350$\\
\vspace{0.2cm}
  $(G_T^{S1},N_B^{S1})=(0.4,1.7) \hspace{0.7cm}$ & $(G_T^S,N_B^S)=(6\cdot 10^{-3},346)$\\
\vspace{0.2cm}
  $(G_T^U,N_B^U)=(0.4,107.32)$ & \\
\vspace{0.2cm}
  $(G_T^{S2},N_B^{S2})=(0.4,\infty)$ & \\
\hline

\end{tabular}
%\end{center}

\caption{Examples of stable and unstable channels for CRDSA (2 replicas) with $N_s=100$ and $I_{max}=20$. \textit{Stable equilibrium points} are marked with a black dot.}
\label{All_channels}
\end{figure}
%%%%%%%%%%

\subsection{\added{Definition of stability}}

Figure~\ref{All_channels} shows the equilibrium contour and the channel load line for various cases. The equilibrium contour divides the ($N_B$,$G_T$) plane in two parts and each channel load line can have one or more intersections with the equilibrium contour. These intersections are referred to as equilibrium points since $G_{OUT}=G_T$. The rest of the points of the channel load line belong to one of two sets: those on the left of the equilibrium contour represent points for which $G_{OUT}>G_T$, thus situations that yield to decrease of the backlogged population; those on the right represent points for which $G_{OUT}<G_T$, thus situations that yield to growth of the backlogged population.  

Therefore we can gather that an intersection point where the channel load line enters the left part for increasing backlogged population corresponds to a \textit{stable equilibrium point}. In particular, if the intersection is the only one, the point is a \textit{globally stable equilibrium point} (indicated as $G_T^G$,$N_B^G$) while if more than one intersection is present, it is a \textit{locally stable equilibrium point} (indicated as $G_T^S$,$N_B^S$). If an intersection point enters the right part for increasing backlogged population, it is said to be an \textit{unstable equilibrium point} (indicated as $G_T^U$,$N_B^U$).

\subsection{\added{Channel state definition}}

Figure~\ref{st} shows a stable channel. The globally stable equilibrium point can be referred as \textit{channel operating point} in the sense that we expect the channel to operate around that point. With the word around we mean that due to statistical fluctuations, the actual $G_T$ and $G_B$ (and thus also $G_{IN}$ and $G_{OUT}$) may differ from the expected values. In fact, the actual values have binomially distributed probability for $G_B$ and $G_T$ with finite M
%
%\begin{equation}
%Pr\bigg\{G_B=\frac{b}{N_s}\bigg\}=\frac{\binom{N_B}{b}\cdot p_r^b\cdot (1-p_r)^{N_B-b}}{N_s}
%\end{equation}
%
%\begin{equation}
%Pr\bigg\{G_T=\frac{t}{N_s}\bigg\}=\frac{\binom{M-N_B}{t}\cdot p_0^t\cdot (1-p_0)^{N_B-t}}{N_s}
%\end{equation}
%
and Poisson distributed probability for $G_T$ with infinite $M$.
%
%\begin{equation}
%Pr\bigg\{G_T=\frac{t}{N_s}\bigg\}=\frac{(\lambda^{t}\cdot e^{-\lambda})/(t!)}{N_s}
%\end{equation}
%
%where $b$ and $t$ are respectively the number of backlogged and thinking users transmitting in a certain frame. 
Nevertheless, the assumption of taking the expected value has been already validated in \cite{stab1}.

Figures~\ref{unstFin} and~\ref{unstInf} show two unstable channels respectively for finite and infinite number of users. Analyzing this two figures for increasing number of backlogged packets, the first equilibrium point is a stable equilibrium point. Therefore the communication will tend to keep around it as for the stable equilibrium point in Figure~\ref{st} and we can refer to it once again as \textit{channel operating point}. However, due to the abovementioned statistical fluctuations, the number of backlogged users could pass the second intersection and start to monotonically increase. 

In the case of finite $M$, this increment goes on till a new intersection point is reached, while in the case of infinite $M$ the expected number of backlogged users increases without any bound. In the former case, this third intersection point is another stable equilibrium point known as \textit{channel saturation point}, so called because it is a condition in which almost any user is in B state and $G_{OUT}$ approaches zero. In the latter case, we can say that a \textit{channel saturation point} is present for $N_B\rightarrow\infty$. 

Finally Figure~\ref{overL} shows the case of an overloaded channel. In this case there is only one equilibrium point corresponding to the channel saturation point. Therefore, even though the channel is nominally stable, the point of stability occurs in a non-desired region. For this reason this case is separated and distinguished from what is intended in this work as \textit{stable channel}. 

\section{Packet Delay Model} \label{sec:delay}
\begin{figure}[tbh!]
\centering
\includegraphics [scale = 0.44] {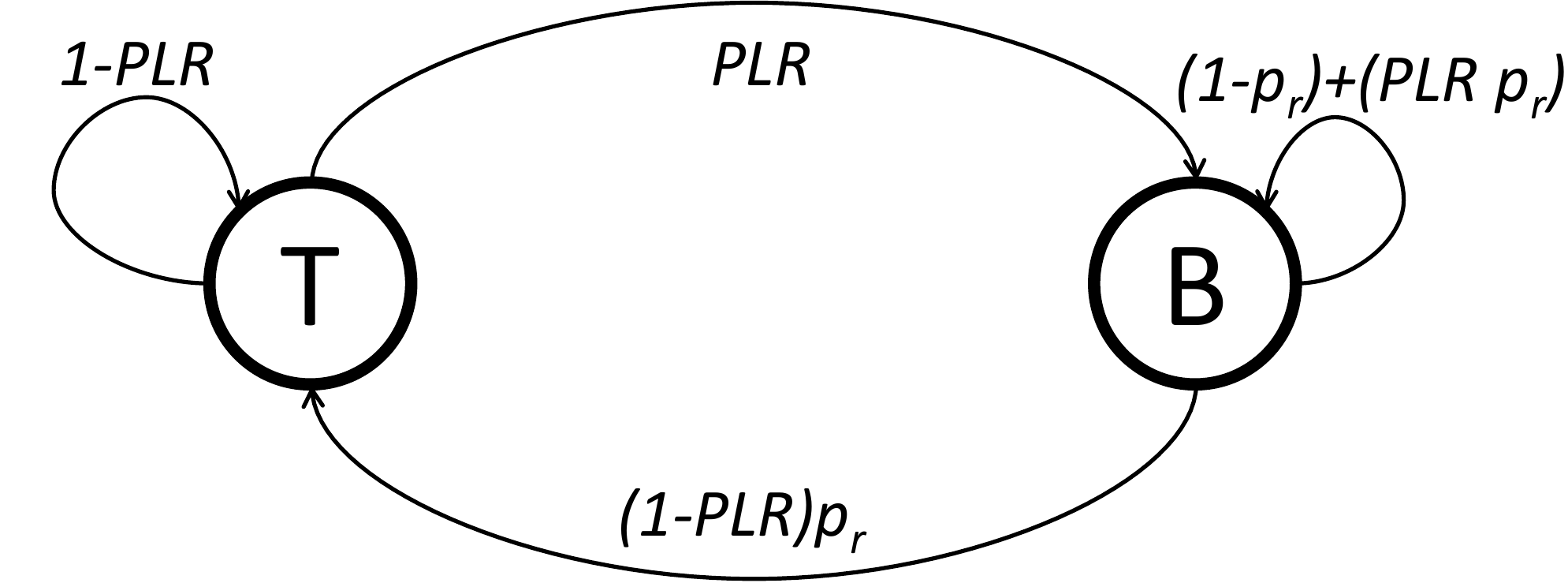}
\caption{Markov Chain for the Packet Delay analysis.}
\label{FSM}
\end{figure}

\replaced{Once we are sure that the channel is working at its}{Assuming a channel at its} \textit{channel operating point}, we would like to know what is the expected distribution and average delay associated to packets that are successfully received at the gateway. This can be described using a discrete-time Markov chain with the two states B and T (Figure~\ref{FSM}). 

The edges emanating from the states represent the state transitions occurring to users which depend on $p_r$ and the expected $PLR$ at the channel operating point. If a packet transmitting for the first time receives a positive acknowledgment the user stays in T state while in case of negative acknowledgment the user switches to B state until successful retransmission has been accomplished. We assume a frame duration to be the discrete time unit of this Markov chain. Therefore the packet delay $D_{pkt}$ is calculated as the number of frames that elapse from the beginning of the frame in which the packet was transmitted for the first time, till the end of the one in which the packet was correctly received.

According to the definition given above, the delay distribution is entirely described by

\begin{equation}\label{FSMeq}
Pr\{D_{pkt}=f\} =
\begin{cases}
1-PLR\text{\,,\ \ \ \ \ \ \ \ \ \ \ \ \ \ \ \ \ \ for  } f=1 \\
\\
PLR\ [p_r\ (1-PLR)] \cdot \\
\cdot [1-p_r+PLR\ p_r]^{f-2}     \text{\ ,\ for  } f>1
\end{cases}
\end{equation}

which results in the average delay

\begin{equation}\label{FSMAvEq}
Av[D_{pkt}]=\sum_{f=1}^{\infty}  f \cdot Pr\{D_{pkt}=f\}
\end{equation}

Equation~\ref{FSMAvEq} can alternatively be rewritten in a simpler and more intuitive form by means of Little's Theorem \cite{stab2}\cite{Little} considering that the average number of backlogged users in a stable system is equal to the average time spent in backlogged state, multiplied by the arrival rate of new packets $G_T$ (that we know to be equal to $G_{OUT}$ at the operational point). Therefore

\begin{equation}\label{LittleFor}
Av[D_{pkt}]=\frac{N_B}{G_{OUT}\cdot N_s}
\end{equation}

where the presence of $N_s$ in the formula has the aim of normalizing the delay to the frame unit.

\begin{table}[tbh!]
\centering
\begin{tabular}{c c c c c c c}
    \hline\\
    $p_r$ & $Av[D_{pkt}]_{ana}$ & $Av[D_{pkt}]_{sim}$ & $PLR$  \\ \\ 
    \hline \hline
    1 & 543.48 & 541.89 & 0.99816 \\ 
    \hline
    0.4 & 3.17 & 3.16 & 0.465 \\ 
    \hline
    0.2 & 2.01 & 2.12 & 0.168 \\ 
    \hline
    0.05 & 2.53 & 2.66 & 0.0713 \\ 
    \hline
    0.005 & 5.08 & 5.22 & 0.02 \\ 
    \hline
  \end{tabular}

\caption{Analytic and simulated average packet delay at the channel operating point for CRDSA with $N_s=100 \ slots$, $I_{max}=20$, $M=350$ and $p_0=0.18$.}
\label{tab:delayres}
\end{table}

\subsection{\added{Model validation and retransmission probability's effect}}

\begin{figure}[tbh!]
\centering
\includegraphics [scale = 0.4] {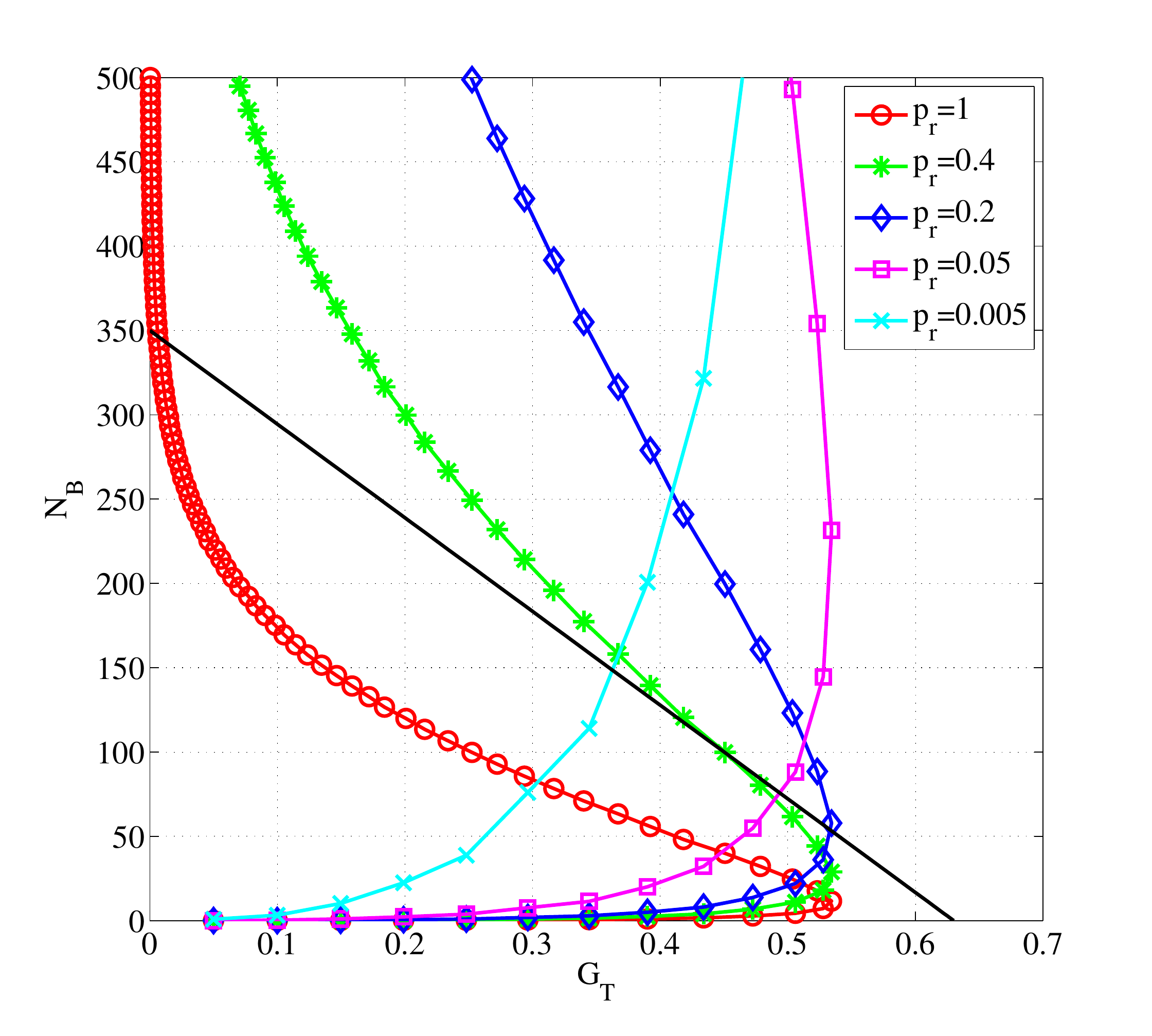}
\caption{Equilibrium curves for CRDSA with $N_s=100 \ slots$, $I_{max}=20$, $M=350$ and $p_0=0.18$.}
\label{eq_curve_delay}
\end{figure}

\begin{figure}[tbh!]
\centering
\includegraphics [scale = 0.4] {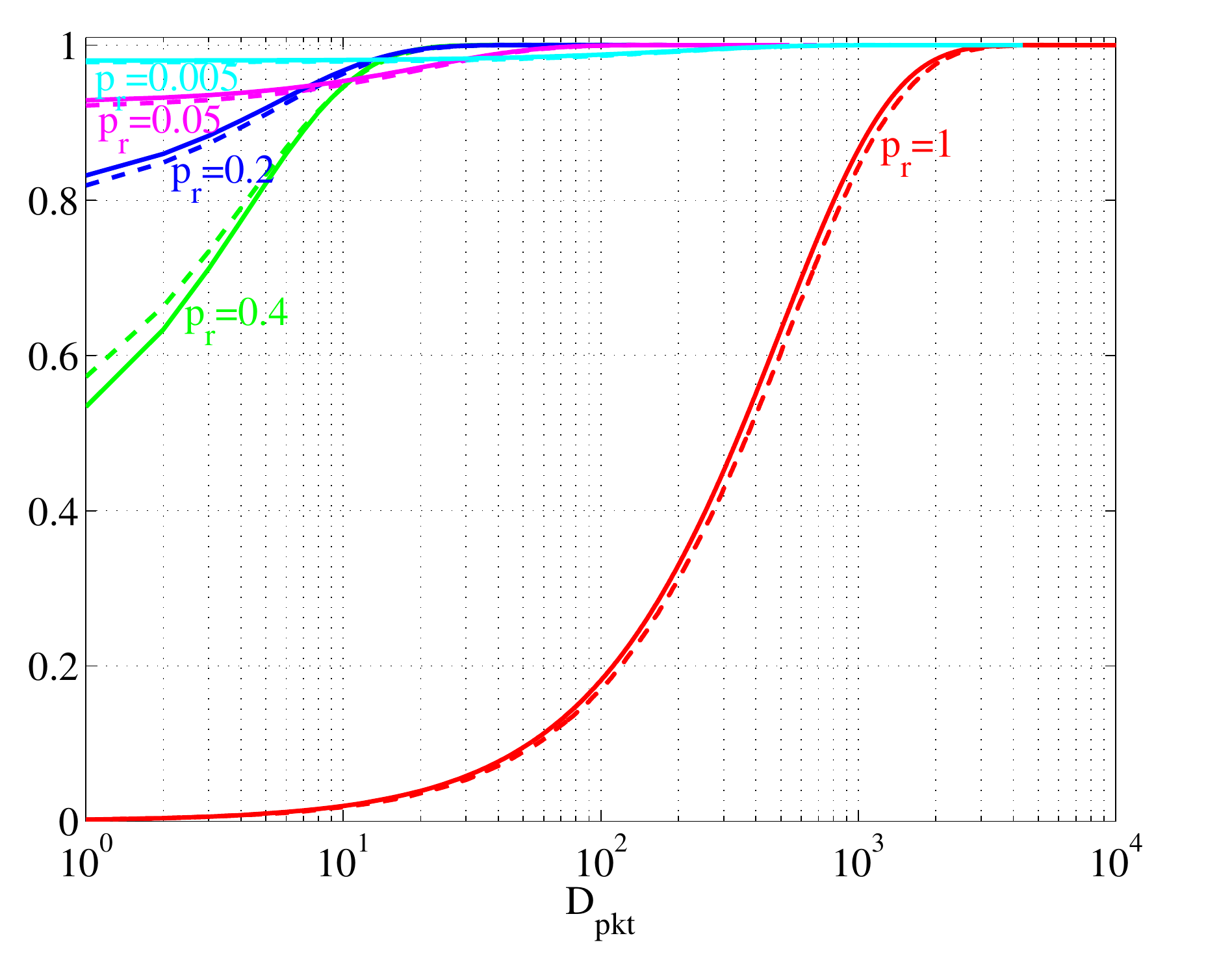}
\caption{Packet delay cumulative distribution for CRDSA with $N_s=100 \ slots$, $I_{max}=20$, $M=350$ and $p_0=0.18$.}
\label{cumCurve}
\end{figure}

Figures~\ref{eq_curve_delay} and~\ref{cumCurve} show some results based on the example of overloaded channel given in Section III\added{, in order to validate the model and demonstrate the effect of $p_r$ on the packet delay}. In particular $p_r$ is progressively decreased in order to evaluate how distribution and average packet delay change. Figure~\ref{eq_curve_delay} represents the channel load line and equilibrium curves for several $p_r$ values. 

When $p_r$ is decreased, the equilibrium contour moves upwards and the point of equilibrium is found for bigger values of throughput while $N_B$ decreases. Therefore, from Little's theorem, it is intuitive to expect that users will spend less time in B state. If we keep decreasing $p_r$ after the throughput peak is reached, the throughput starts to decrease while the number of backlogged users increases again thus yielding to an increase of the average packet delay. This is confirmed by the results given in Table~\ref{tab:delayres}.

Figure~\ref{cumCurve} shows that while having a small $p_r$ is beneficial for the $PLR$, it also means that users in B state will generally wait longer before retransmitting a packet. As a consequence, the smaller the $p_r$ value the less steep the cumulative distribution curve is, so that at a certain point curves with bigger $p_r$ overcome it in terms of probability to receive a packet before a certain time deadline. Finally the simulated average packet delay associated to the examples in Figure~\ref{cumCurve} is reported in Table~\ref{tab:delayres} together with the expected results, demonstrating the validity of the model.

\section{First Exit Time}\label{FET}
In Section III the concept of unstable channel has been introduced.\added{ Beside knowing that a certain channel is unstable, it is also of big interest to quantify how much the channel is unstable}. The First Exit Time (FET) \deleted{allows us to quantify how much the channel is unstable in terms of}\added{calculates the} time \deleted{needed to exit}\added{spent around the operational point by an unstable channel before leaving} the stability region. To calculate this parameter we consider a discrete-time Markov chain which uses the number of backlogged users $N_B$ as describing state, since equilibrium points and more generally the state of the communication can be uniquely identified with it. 

\subsection{\added{Transition matrix definition}}

Let us define $P$ as the transition matrix for this Markov chain, with each element $p_{ij}$ denoting the probability to pass from a number of backlogged users $N_B=j$ to a number of backlogged users $N_B=i$. To calculate each $p_{ij}$ value we need to know the probability that having a certain number of backlogged users $j$ and a certain number of thinking users $M-j$, respectively $b$ backlogged users and $t$ thinking users have transmitted and $s=t-(i-j)$ out of $t+b$ packets have been correctly received. Therefore

\begin{equation} \begin{split} \label{finiteM_trans}
p_{ij}= \sum_{t=0}^{M-j} \sum_{b=0}^{j} \binom{M-j}{t} p_0^t (1-p_0)^{M-j-t}\cdot \\
\cdot \binom{j}{b} p_r^b (1-p_r)^{j-b}\cdot q(s|t+b)
\end{split} \end{equation}

where $q(s|t+b)$ is the probability that $s$ packets are correctly decoded  if $t+b$ packets are transmitted\footnote{
For the same reason claimed \replaced{in footnote 3}{before} for the PLR\deleted{ values used for the stability model}, also $q(s|t+b)$ values are calculated from simulations over a big amount of runs (over $10^6$ per $(t+b)$ value) since analytical formulas able to tightly describe the probability for a packet to be correctly decoded have not been found yet. Nevertheless in the followings a method to reduce this computational effort is introduced.}. 
Let us then call $Q$ the matrix with elements $q_{s,t+b}$ corresponding to probability $q(s|t+b)$. 
 \begin{figure}[t!]
\centering
\includegraphics [scale = 0.35] {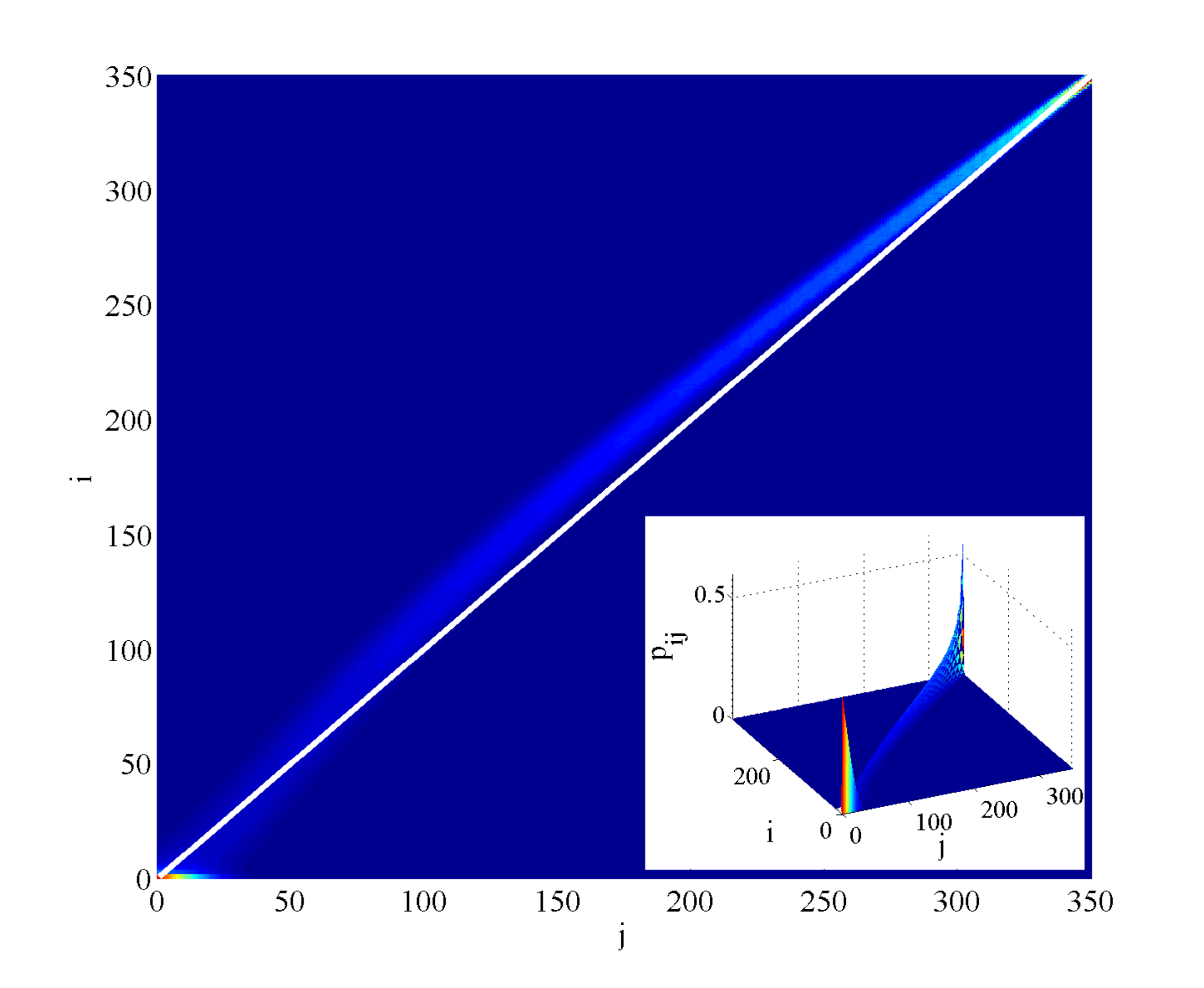}
\caption{P matrix for CRDSA with $N_s=100$, $I_{max}=20$, $M=350$, $p_0=0.143$, $p_r=1$.}
\label{P_fin}
\end{figure}

In Figure~\ref{P_fin}\added{ an example of} the resulting P matrix is shown. Accordingly to the bidimensional representation used in Section~III, we can see here that there are two peaks corresponding to the two expected locally stable equilibrium points. However, it is more fruitful for our considerations to look at this graph from an $(i,j)$ plane perspective plotting a line for $i=j$. Points below this line represent decrements of the number of backlogged packets while points above represent increments. Non-zero probability points look like a wake close to the line $i=j$, highlighting that possible transitions are found only for small changes in the number of backlogged packets compared to the total population $M$. Taking single values of $j$, we can roughly identify 3 regions of the wake:
\begin{itemize}
\item for small values of $j$ (close to the channel operating point) non-zero probabilities are found both for increase and decrease of $N_B$;
\item for values of $j$ close to $M$ (i.e. close to the channel saturation point) non-zero probabilities are found both for increase and decrease of $N_B$;
\item for the rest of the $j$ values non-zero probabilities are almost exclusively found for $i>j$ so that the number of backlogged users monotonically increases.
\end{itemize}

\added{Therefore in this particular case the system has probability zero of returning to the stability region after $N_B$ crosses 120. We want to highlight here that the presence of a \textit{no-return} point is scenario-dependent in the sense that for a different set of values $M$, $p_0$ , $p_r$ , $N_s$ it is not necessarily true that the probability of coming back from a certain value $N_B$ is zero: sometimes could be negligible, sometimes could not, even though for the typical settings used in these scenarios the last case is really rare. Nevertheless this highlights one of the advantages of calculating the P matrix: it tells us theoretically if there is any probability of coming back to the region of stability and in the case the answer is no, tells us what is the \textit{no-return} value of $N_B$.}

\subsection{\added{State probability matrix and FET calculation}}

Based on the $P$ matrix, it is now possible to calculate the probability $B^{f+1}_i$ to be in a certain state $N_B=i$ at frame $f+1$, given a certain probability distribution $B^f=[B^{f}_0\ B^{f}_1\  \cdots \ B^{f}_M]$ at frame $f$:

\begin{equation}\label{state_trans}
%[
    \begin{bmatrix}
       B^{f+1}_0 \\[0.3em]
       B^{f+1}_1 \\[0.3em]
       \vdots \\
       B^{f+1}_M \\[0.3em]
     \end{bmatrix}
=
    \begin{bmatrix}
       p_{0,0} & \cdots & p_{0,M} \\[0.3em]
       p_{1,0} & 	  & \vdots  \\[0.3em]
       \vdots  & \ddots & \vdots  \\
       p_{M,0} & \cdots & p_{M,M}\\[0.3em]
     \end{bmatrix}
\cdot
    \begin{bmatrix}
       B^f_0 \\[0.3em]
       B^f_1 \\[0.3em]
       \vdots \\
       B^f_M \\[0.3em]
     \end{bmatrix}
%]
\end{equation}

Generalizing Equation~\ref{state_trans}, the state distribution at each frame $(f+1)$ can always be calculated from the initial probability distribution $B^0=[B^{0}_0\ B^{1}_0\  \cdots \ B^{M}_0]$ and the matrix $P^{f+1}$. \deleted{Therefore}\added{Given the state probability matrix}, it is now possible to calculate the average and cumulative distribution of the FET from a given initial state. In this example we consider as initial state $B^0=[\ 1\ 0\ 0\ \cdots \ 0\ ]$\added{ that is the case in which the communication starts with zero backlogged users}.

\begin{figure}[t!]
\centering
\includegraphics [scale = 0.37] {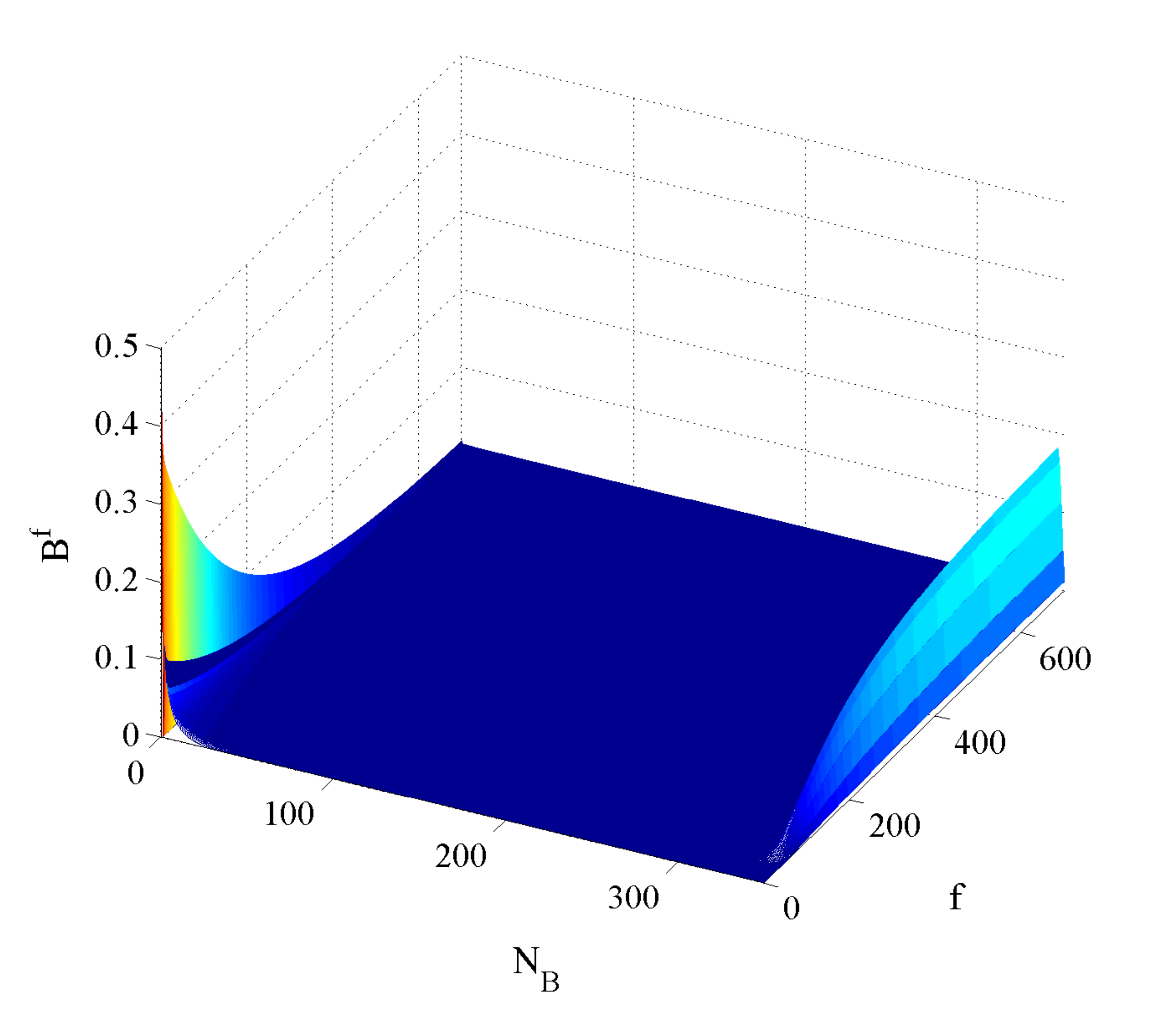}
\caption{Evolution of $B^{f}$ for the P matrix in figure~\ref{P_fin}.}
\label{NBevo}
\end{figure}

Figure~\ref{NBevo} reports the evolution frame after frame of the state probability for the P illustrated in Figure~\ref{P_fin}. As we can see, when $f$ increases the probability of being around the locally stable point in the region of high throughput decreases while the probability of being in saturation increases.

\subsection{\added{Reduced transition matrix}}

The entire description of any state of this Markov chain is of great interest but most of the times impractical, mainly because the dimension of the $P$ matrix is equal to $M^2$\deleted{ and P is derived from simulations}. Therefore it could take a huge computational effort to calculate it. Nevertheless it is of interest to know the probability that the stability region around the operational point has been left in a certain moment, regardless of the actual state once we enter the region that yields to saturation. 

In fact, it has been already pointed out in the description of Figure~\ref{P_fin} that the further the communication gets from the first point of local stability, the smaller is the probability that the communication will come back to the stability region. For this reason, as a first approximation, it is possible to consider an absorbing state $N_B^{abs}=\lceil N_B^U\rceil+1$ that groups all the probabilities for states from $\lceil N_B^U\rceil+1$ to $M$. Doing so, our $P$ matrix can be rewritten as 

\begin{equation}\label{P_rid}
P =
    \begin{bmatrix}
       p_{0,0} & \cdots & p_{0,N_B^U} & 0 \\[0.6em]
       \vdots   & \ddots & \vdots & \vdots \\[0.6em]
       p_{N_B^U,0} & \cdots & p_{1,N_B^U} & 0 \\[0.9em]
       1-(\sum_{\alpha=0}^{N_B^U} p_{\alpha,0}) & \cdots & 1-(\sum_{\alpha=0}^{N_B^U} p_{\alpha,N_B^U}) & 1\\[0.6em]
     \end{bmatrix}
\end{equation}

The $P$ matrix in Equation~\ref{P_rid} takes much less effort than calculating the entire $P$ matrix since only $(\lceil N_B^U\rceil+1)^2$ elements need to be computed thus speeding up calculations. Considering that $\sum_{\alpha=0}^{N_B^{U}}p_{\alpha,j}$ must be equal to $1$, the probability to get to the absorbing state $N_B^{abs}$ from a certain state $j$ can be computed as  $p_{i,j}=1-(\sum_{\alpha=0}^{N_B^{U}} p_{\alpha,j})$.
Moreover, since $N_B^{abs}$ is an absorbing state, any probability to leave this state is zero while the probability of staying in the absorbing state is equal to $1$.

\begin{figure}[t!]
\centering
\includegraphics [scale= 0.36] {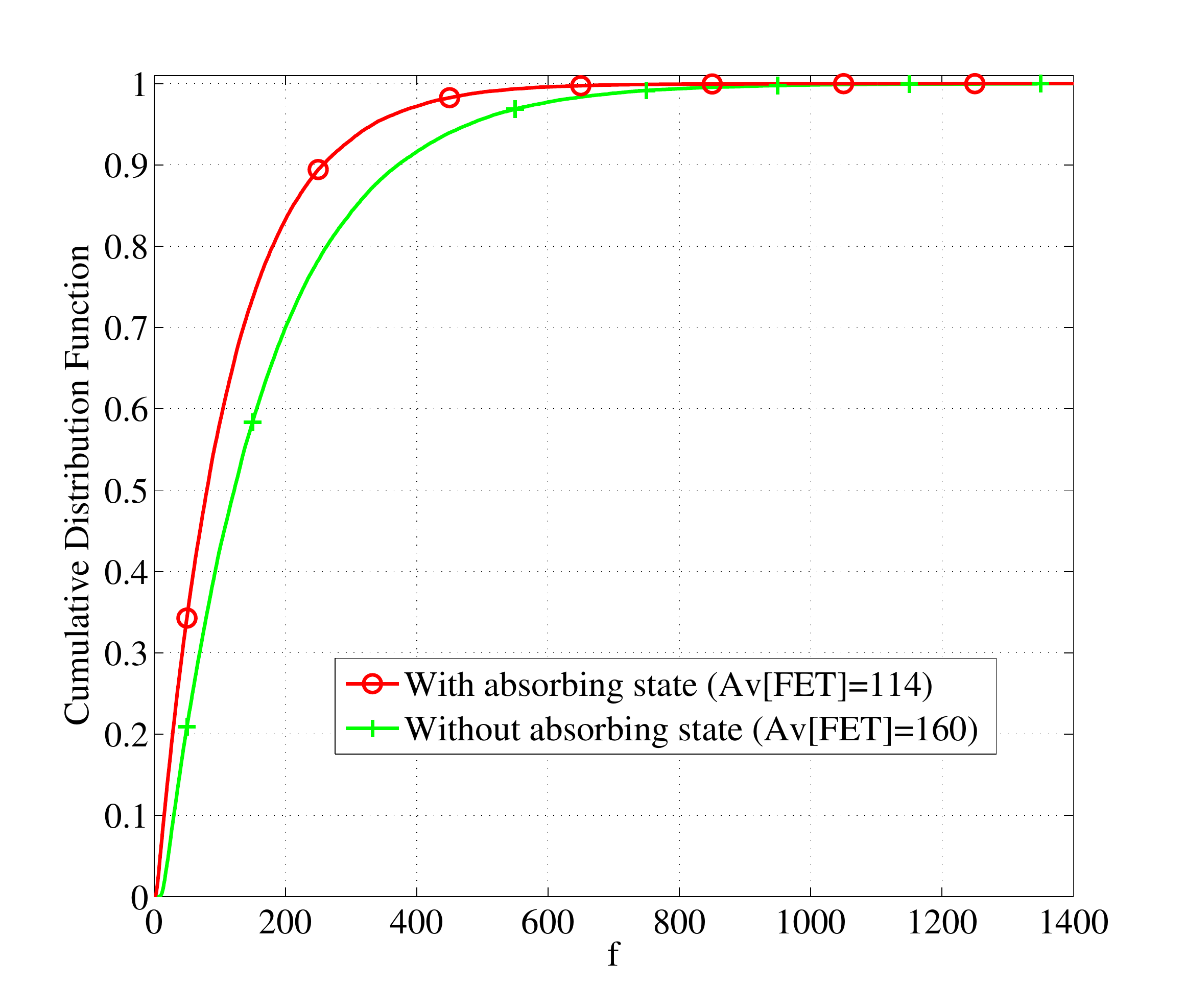}
\caption{FET cumulative distribution for CRDSA with $N_s=100$, $I_{max}=20$, $M=350$, $p_0=0.18$, $p_r=1$.}
\label{FETcdf}
\end{figure}

\subsection{\added{Absorbing state validity}}

\added{Since reducing the transition matrix by means of an absorbing state is an approximation, we}\deleted{We} now aim at verifying the validity of this approximation. Figure~\ref{FETcdf} shows the distribution for the \deleted{analytic }results \replaced{obtained with the reduced matrix}{with absorbing state} and the simulated results\deleted{ without the approximation of the absorbing state}. \replaced{R}{In particular, r}egarding simulations, if $N_B$ crosses the value $\lceil N_B^U\rceil$ but it comes back to the stability region, the FET is reset and the next exit is waited to set it again. For the analytic results the FET is calculated accordingly to the reduced matrix in Equation~\ref{P_rid}. 

\begin{figure}[t!]
\centering
\includegraphics [scale= 0.34] {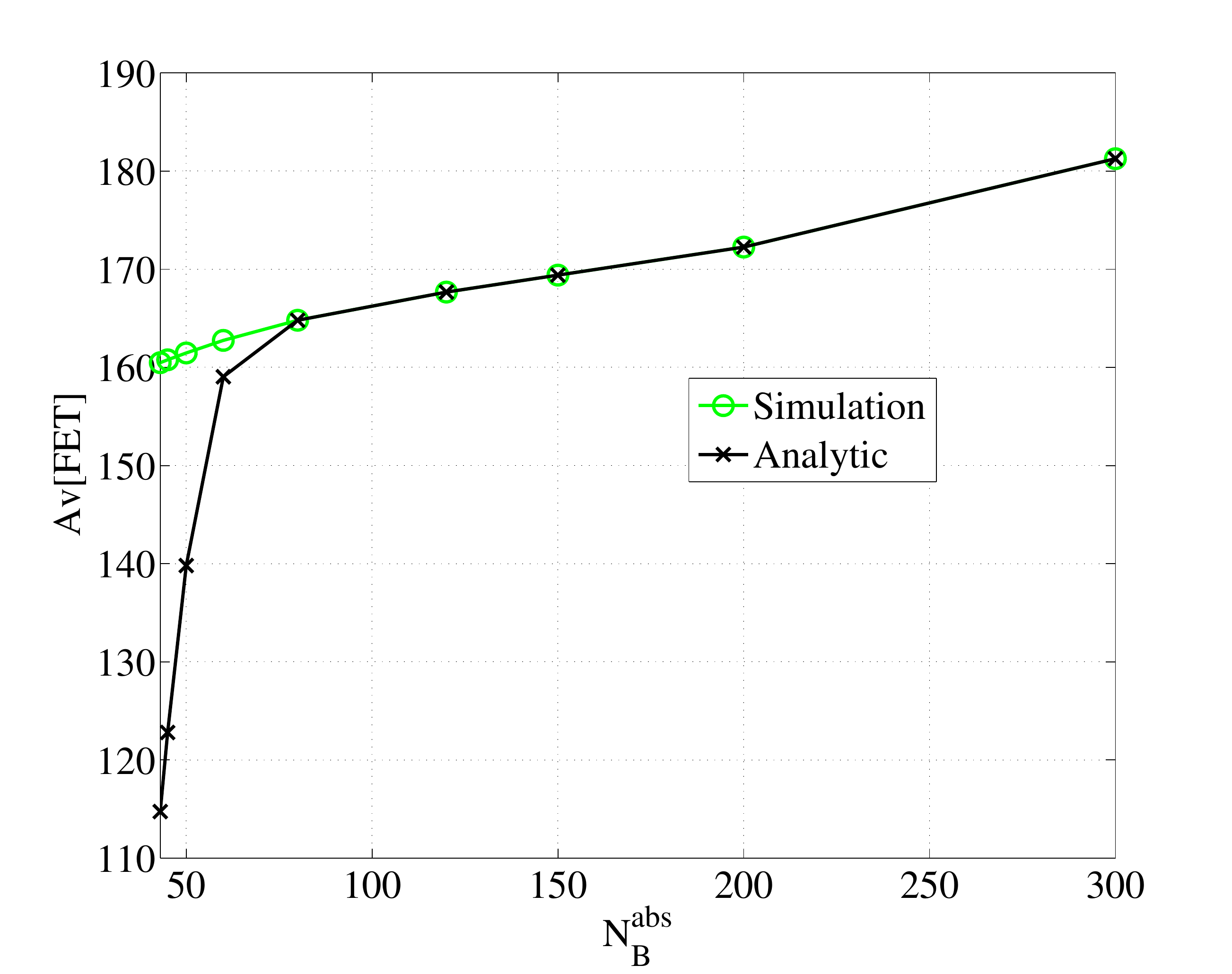}
\caption{Simulated and analytic average FET when increasing $N_B^{abs}$ for CRDSA with $N_s=100$, $I_{max}=20$, $M=350$, $p_0=0.18$, $p_r=1$.}
\label{Delta}
\end{figure}

As already pointed out in the description of Figure~\ref{P_fin}, when the number of backlogged packets $N_B$ is still close to the unstable equilibrium point, a certain probability that the communication will come back from the instability region is present. This is the reason why in Figure~\ref{FETcdf} the two curves \replaced{show}{present} a noticeable difference. However, this problem can be solved by considering $N_B^{abs}=\lceil N_B^U\rceil+1+\Delta$ as absorbing state, where $\Delta$ is a positive integer big enough to ensure that the probability of false exits from the stability region (i.e. the probability that $N_B$ crosses the stability region but comes back to values $< N_B^{abs}$) is sufficiently low. Doing so, simulation and analytical results with the approximation of absorbing state perfectly match. Figure~\ref{Delta} shows how the difference among the two average FETs decreases while increasing the chosen value for $N_B^{abs}$. 

In conclusion, we can state that analytic results with the approximation of absorbing state represent a lower bound for the actual FET. Therefore during this kind of analysis a tradeoff between tightness of the results and simplicity of computation is present.

\subsection{\added{Reduced transition matrix for infinite population}}

\replaced{T}{Finally, we claim that t}he model presented is also adaptable to the case of infinite population. In fact the probability transition matrix $P$ can be adapted by considering

\begin{equation} \begin{split} \label{finiteM_trans}
p_{ij}= \sum_{t=0}^{\infty} \sum_{b=0}^{j} \frac{\lambda^t e^{-\lambda}}{t!}\cdot \binom{j}{b} p_r^b (1-p_r)^{N_B-b}\cdot q(s|t+b)
\end{split} \end{equation}

Although in this case $P$ has infinite size, it has been just shown that is possible and practical to reduce the $P$ matrix by means of an absorbing state. With this expedient, also this case can be treaten since $P$ is transformed to an equivalent matrix with finite size.

\section{Design settings}

In this section we will resume the role of some crucial parameters such as $M$, $p_0$, $p_r$ and explain how the communication state changes when changing them and what are the relations to be considered when designing DVB-RCS2 communications in RA mode. In fact, since optimization depends on \textit{constraints} and \textit{degrees of freedom} of the particular study case, a deep investigation of the relation among these crucial parameters is needed in order to understand tradeoffs that have to be faced. Moreover, the same discussion is introductory to understand how control policies can help to obtain better performance and a stable channel even in case of statistical fluctuations that could yield to instability. 

\begin{figure}[t!]
\subfigure [Changing $p_0$] {\label{changingP0} \includegraphics [ scale = 0.24 ]{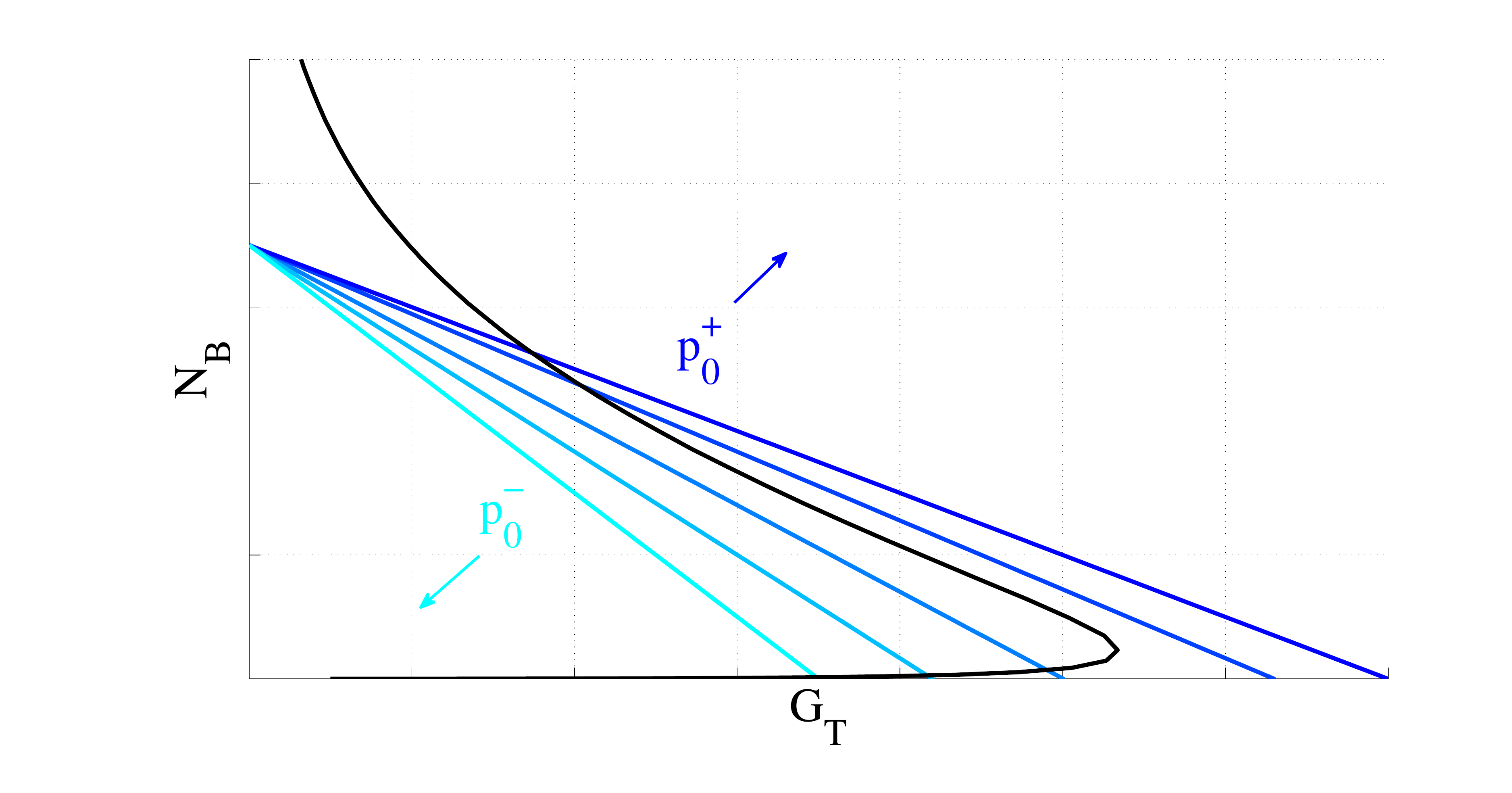}
\label{changingP0}
} \qquad
\subfigure [Changing $M$] {\label{changingM} \includegraphics [ scale = 0.24 ]{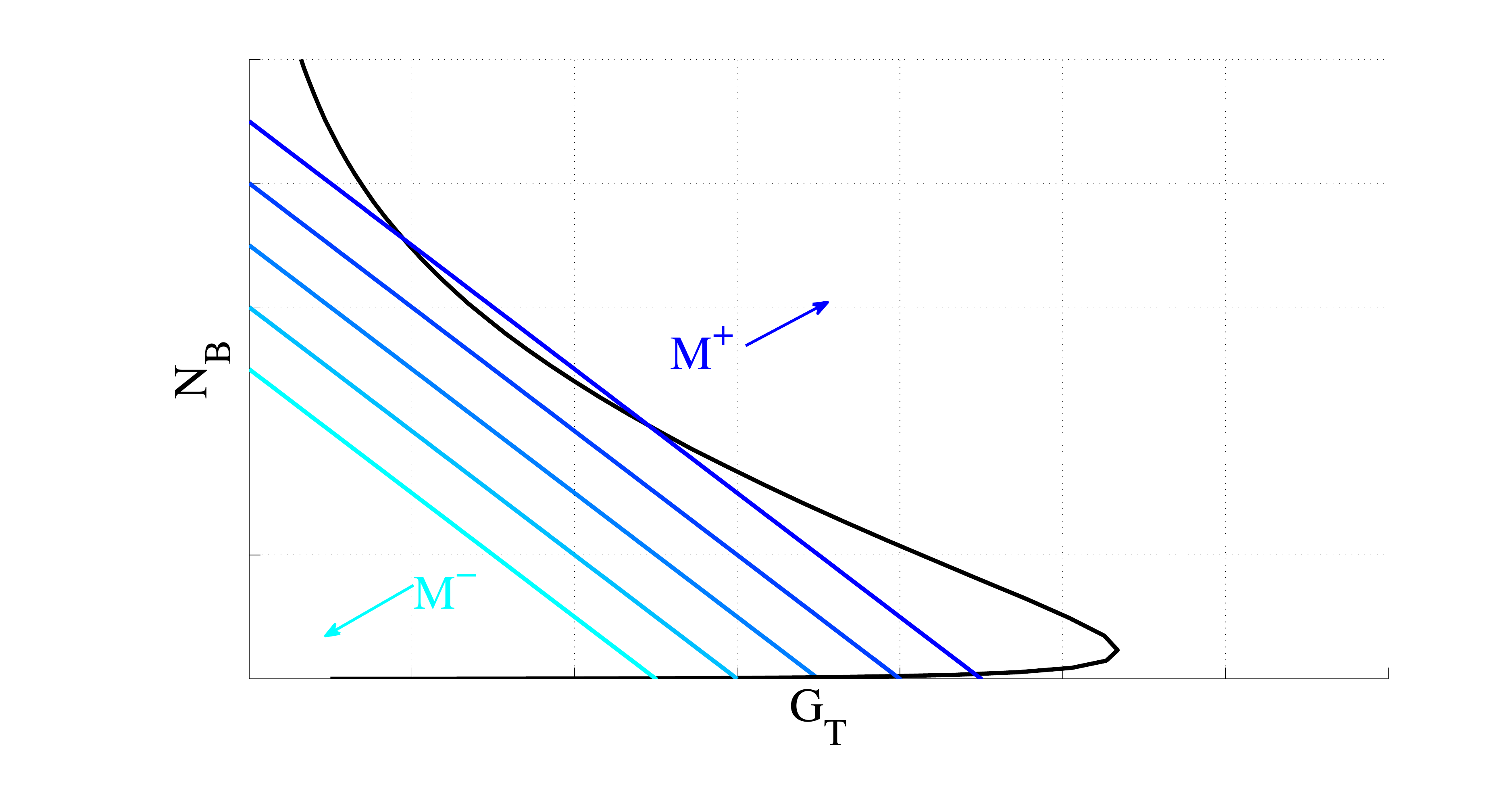}
\label{changingM}
} \qquad
\subfigure [Changing $p_r$] {\label{changingPr} \includegraphics [ scale = 0.24 ]{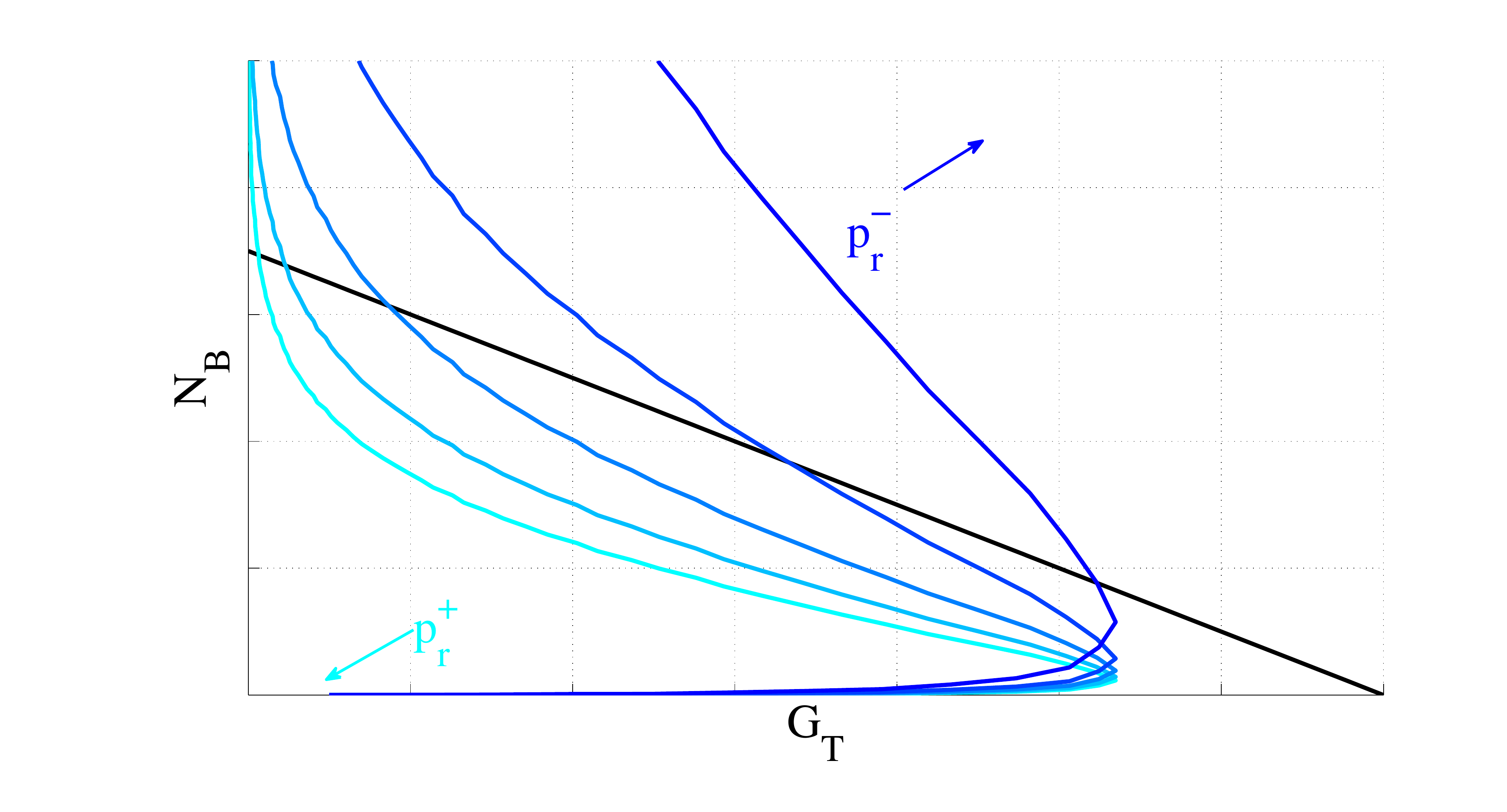}
\label{changingPr}
} \qquad
\\
\\
\caption{Graphical representation of the changes in the ($G_T$,$N_B$) plane when increasing or decreasing one of the parameters' values.}
\label{All_changes}
\end{figure}

\begin{figure}[tbh!]
\subfigure [$p_0$ constraint] {\label{constrP0} \includegraphics [ scale = 0.22 ]{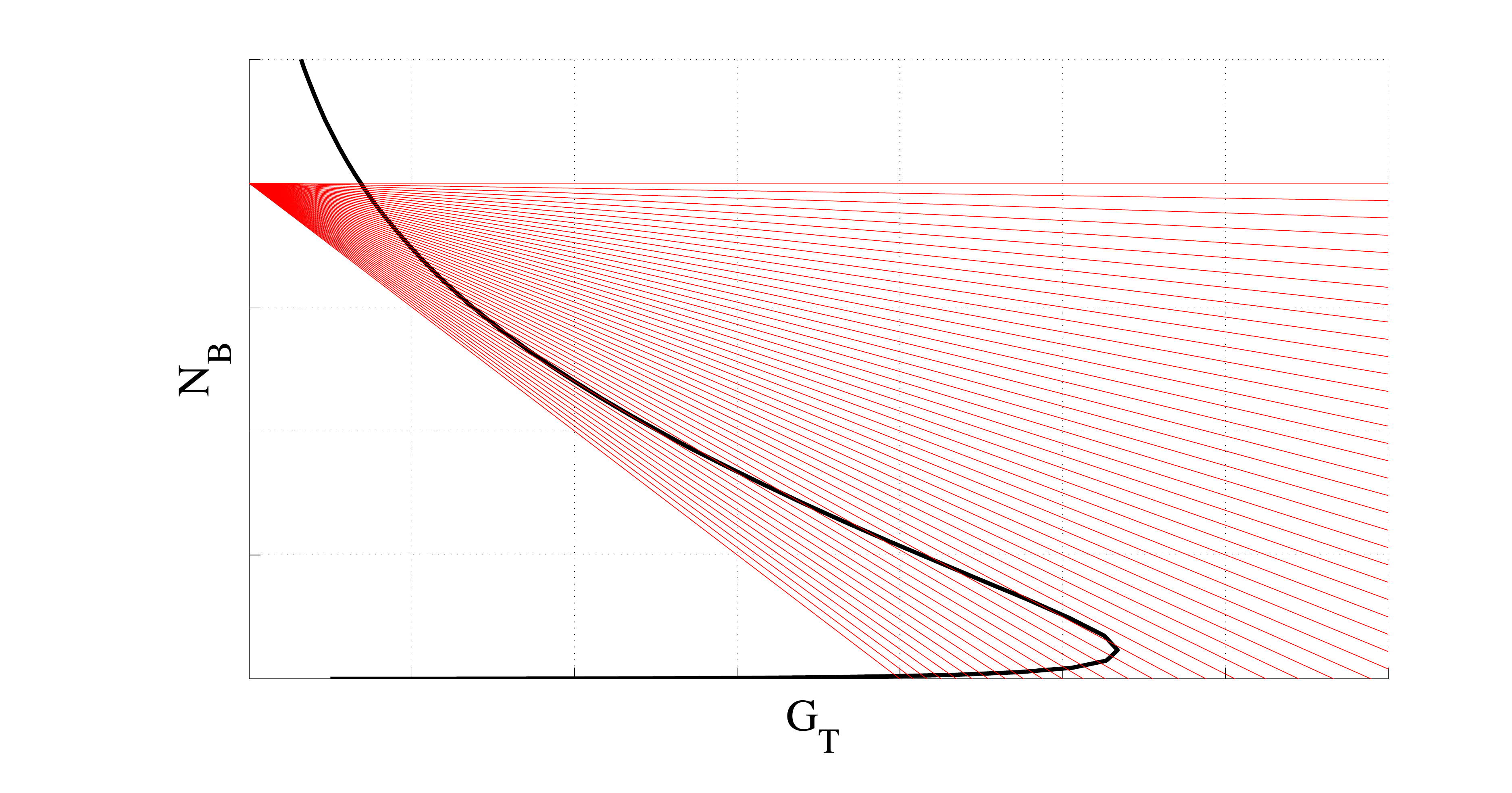}
\label{Admissible area for a certain $p_0^min$ and M fixed}
} \qquad
\subfigure [$M$ constraint] {\label{constrM} \includegraphics [ scale = 0.22 ]{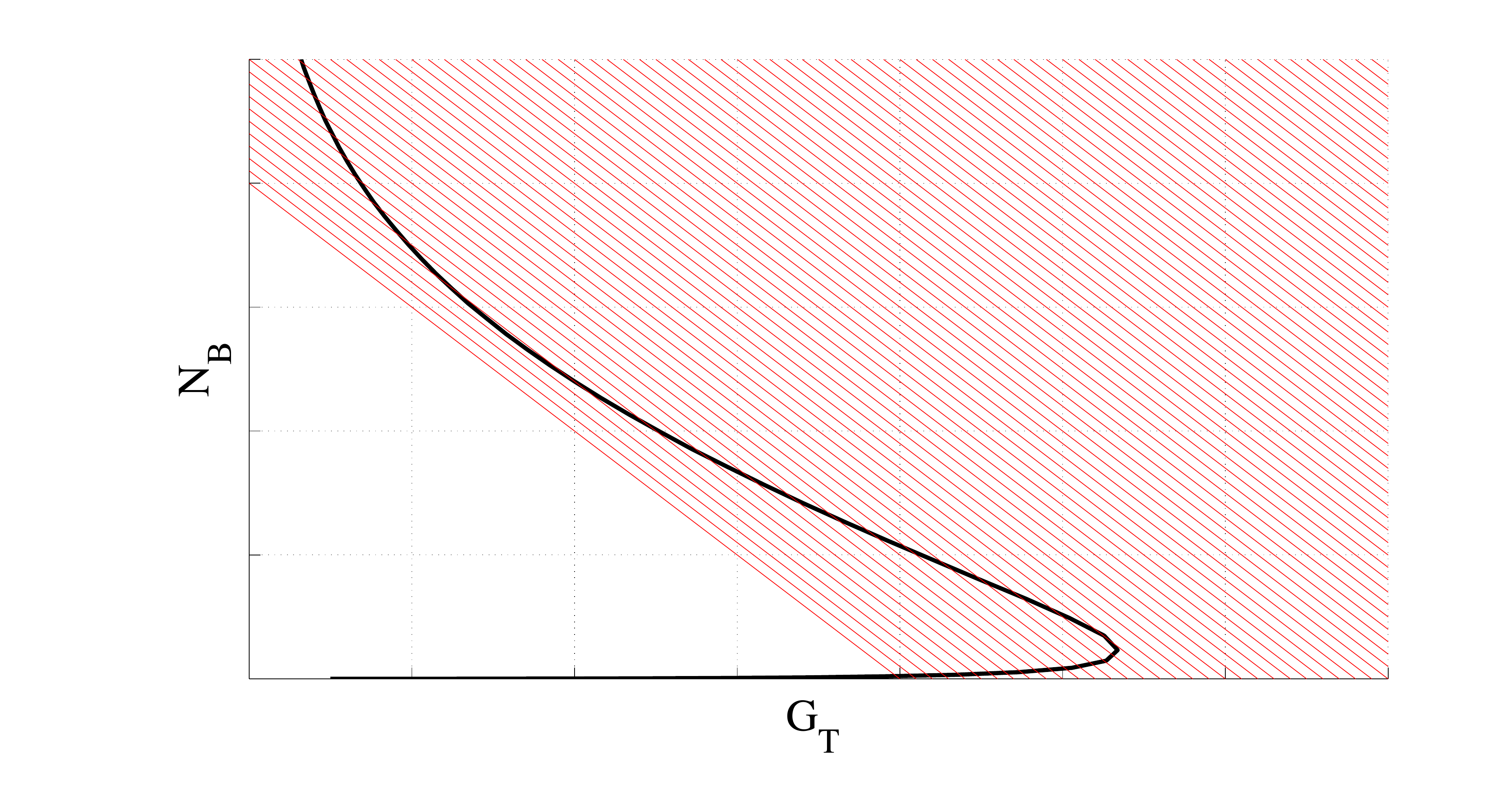}
\label{Admissible area for a certain $M^min$ and p_0 fixed}
} \qquad
\subfigure [Delay constraint] {\label{constrPr} \includegraphics [ scale = 0.22 ]{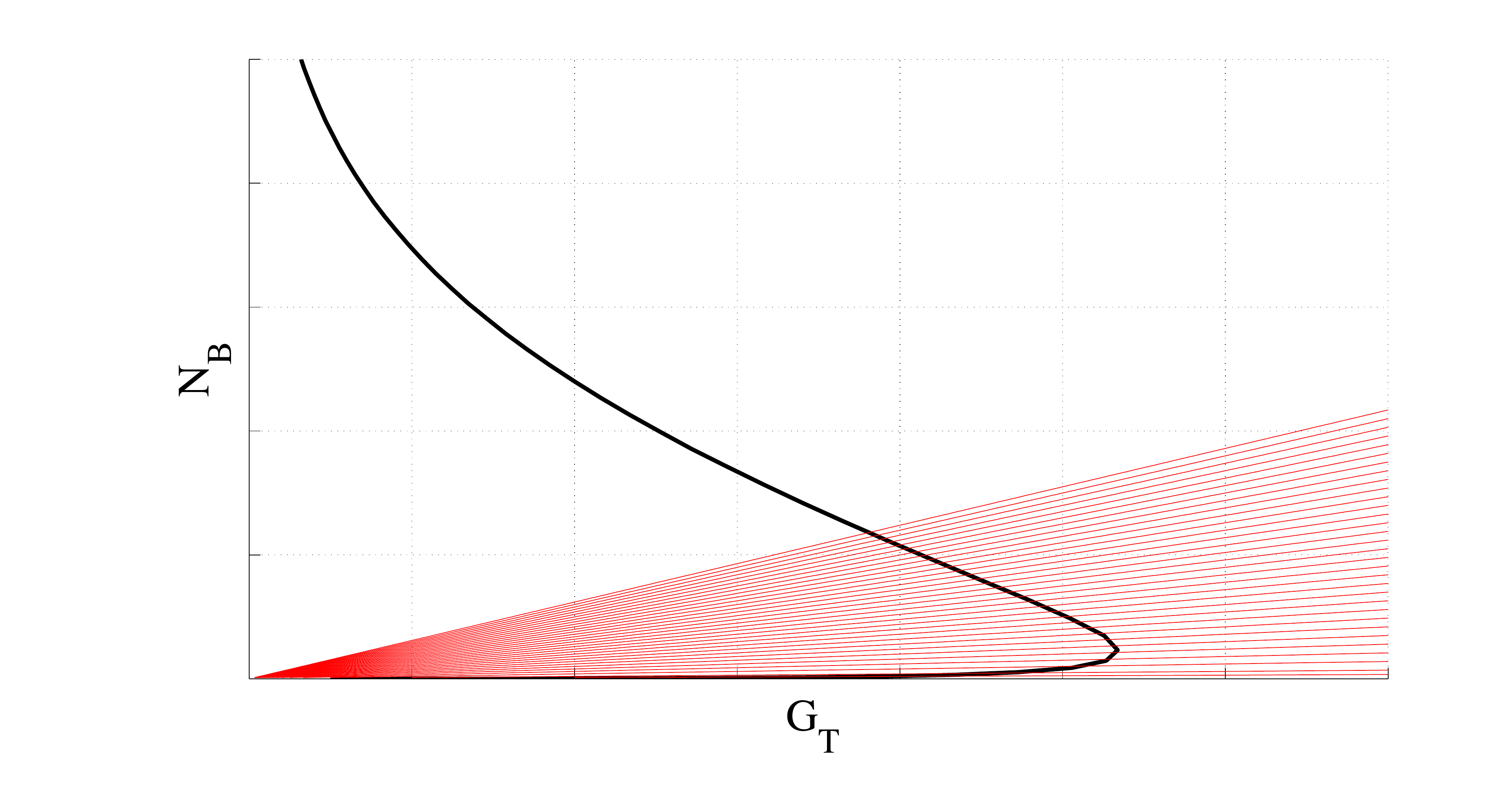}
\label{Admissible area for a certain $Av[D_{pkt}]^max$}
} \qquad
\subfigure [Throughput constraint] {\label{constrPr} \includegraphics [ scale = 0.22 ]{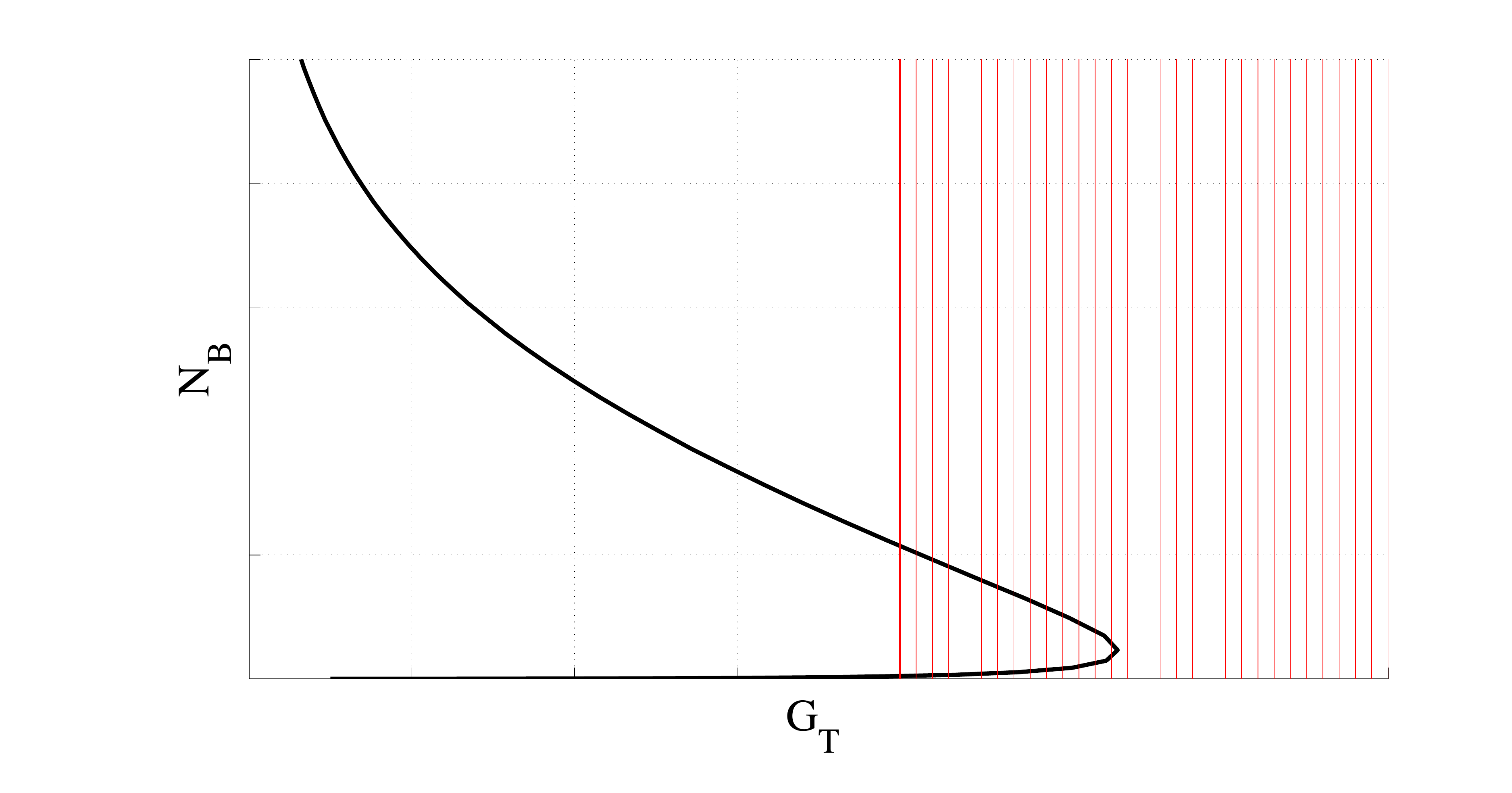}
\label{Admissible area for a certain $G_{OUT}^{min}}
} \qquad
\\
\\
\caption{Graphical representation of the admissible areas for various constraints.}
\label{All_areas}
\end{figure}

\subsection{\added{Degrees of freedom}}

Consider Figure~\ref{All_changes}. The first thing we can notice is that $M$ and $p_0$ only influence the channel load line while changing $p_r$ corresponds to \replaced{modifying}{a change of} the equilibrium contour. In Figure~\ref{changingP0} we can see that $p_0$ influences the slope of the channel load line that becomes steeper when increasing $p_0$. This is intuitive to understand since for fixed $M$, if $p_0^A<p_0^B$ then $G_T^A<G_T^B$. Figure~\ref{changingM} shows that $M$ shifts the channel load line up and down if the population size is respectively increasing or decreasing. From Equation~\ref{LL1} we can notice that the value M is nothing else that what is known in literature as $q$, i.e. the intersection of the line with the y-axis. Finally $p_r$ determines a shift upward of the channel load line for smaller values of $p_r$ (Figure~\ref{changingPr}). As already shown, if the value $p_r$ is sufficiently small, it is possible to stabilize a channel initially unstable. This represents the ground base for one of the control policies shown in the next section.

\subsection{\added{Constraints}}

Concerning possible \textit{constraints}, we can identify the followings

\begin{enumerate}
\item stability
\item maximize $p_0$ or guarantee a value $p_0\geq p_0^{min}$
\item maximize $M$ or guarantee a value $M\geq M^{min}$
\item minimize the delay or guarantee that a certain value $Av[D_{pkt}]^{max}$ is not exceeded
\item maximize the throughput or guarantee a value \ \ \   $G_{OUT}\geq G_{OUT}^{min}$
\end{enumerate}

For constraint 1) it has been previously shown that the stability corresponds to having a single intersection between the channel load line and the equilibrium curve (i.e. a single point of equilibrium)\added{ before the throughput peak}. 
Constraints from 2) to 5) can be graphically represented  on the ($G_T$,$N_B$) plane by the following formulas:

\begin{enumerate}
\setcounter{enumi}{1}
\item \begin{equation}\frac{G_T\cdot N_s}{M-N_B}\geq p_0^{min}\end{equation}
\item \begin{equation}\frac{G_T\cdot N_s}{p_0}+N_B\geq M^{min}\end{equation}
\item \begin{equation}\frac{N_B}{G_T\cdot N_s}\leq Av[D_{pkt}]^{max}\end{equation}
\item \begin{equation}G_{T}\geq G_{OUT}^{min}\end{equation}
\end{enumerate}

In other words, each formula simply represents an admissible area where the operational point has to be, as illustrated in Figure~\ref{All_areas}.

Any study case will be characterized by one or more \textit{degrees of freedom} and one or more \textit{design constraints}. In case of more than one design constraint, the resulting constraint will be the intersection of the respective admissible areas. Therefore we can conclude that the \replaced{designed model}{equilibrium contour} not only \replaced{allows to forecast}{represents the expected} stability and throughput of the communication, but also \added{to easiliy represent} its constraints. This \replaced{enables}{allows} a design that does not need any use of complicated formulas in order to understand feasible settings. 

Notice that the choice of $N_s$ has not been considered as a design parameter. This choice have basically two reasons: first of all in DVB-RCS2, \replaced{the number of slots allocated for RA is dynamically decided by the NCC}{is the NCC that decides how many slots must be allocated for RA}; secondly, the best choice for $N_s$ can always be considered between $100$ and $200$ since a lower value would degrade too much the throughput performance while a bigger value would degrade too much the delay performance.

\section{Packet degree and RA technique}\label{pktdeg}

In the previous sections we have concentrated on the best selection of $p_0$, $p_r$ and $M$ once the number of replicas per packet (packet degree) is chosen. However also the number of packet copies is a design parameter\footnote{In this paper we compare various packet degrees under the assumption of equal received power per replica. For discussion on the power efficiency when considering transponder's average power limits refer to \cite{AvPow}.\added{ Moreover the additional overhead due to the need of more pointers when more copies are sent has been omitted, since it can be considered neglibile with regard to the burst size especially when few copies (no more than 4 or 5) are considered.}}. \added{For this reason, this section analyzes the choice on the number of copies per packet to be sent. Moreover, a comparison on the advantages and disadvantages of CRDSA with regard to SA are outlined.
} 

\begin{figure}[tbh!]
\centering
\includegraphics [width=9 cm] {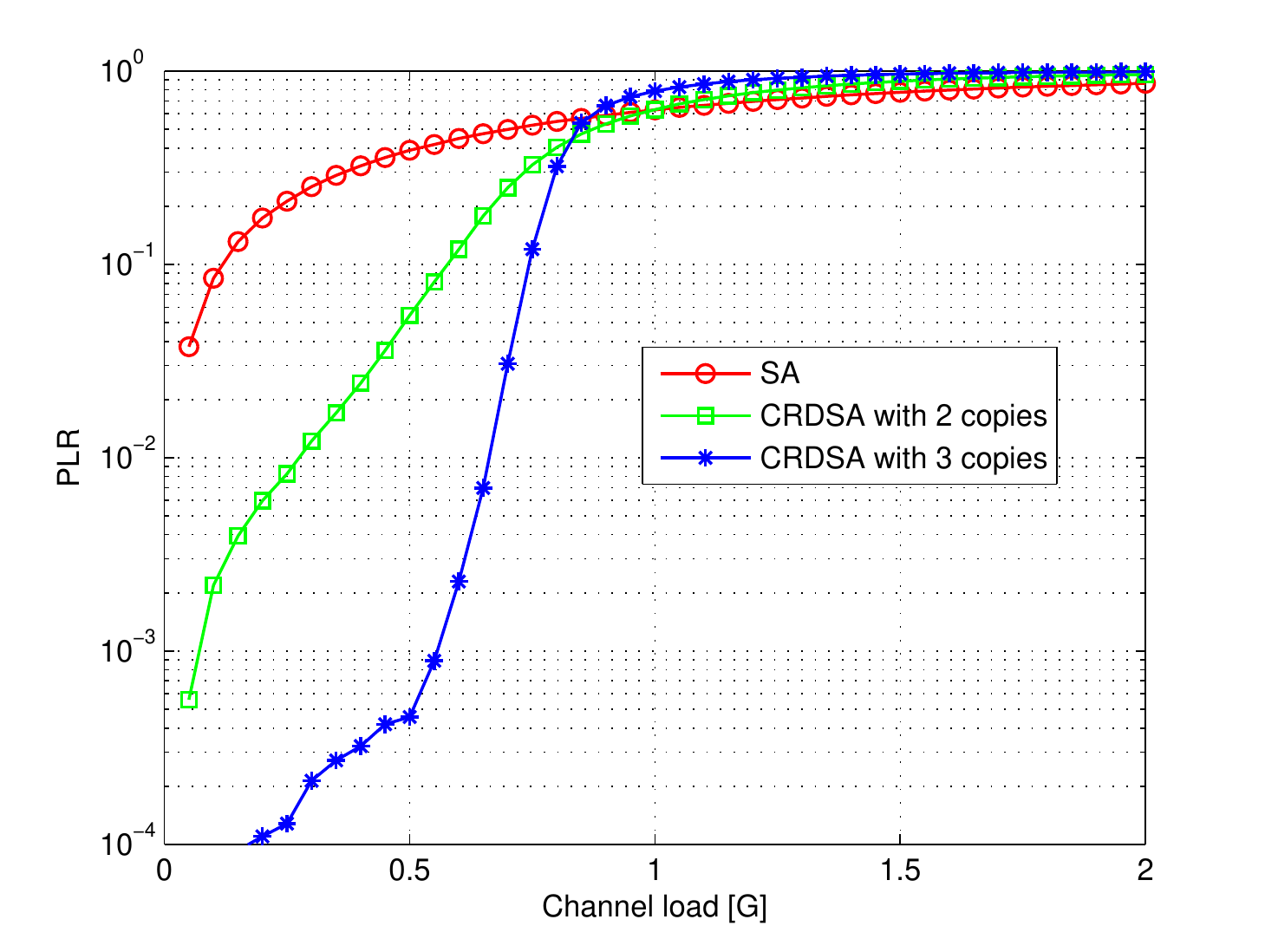}
\caption{Packet Loss Ratio for different packet degrees when $N_S=100$ slots}
\label{plrcomparison}
\end{figure}

\begin{figure}[tbh!]
\centering
\includegraphics [width=9 cm] {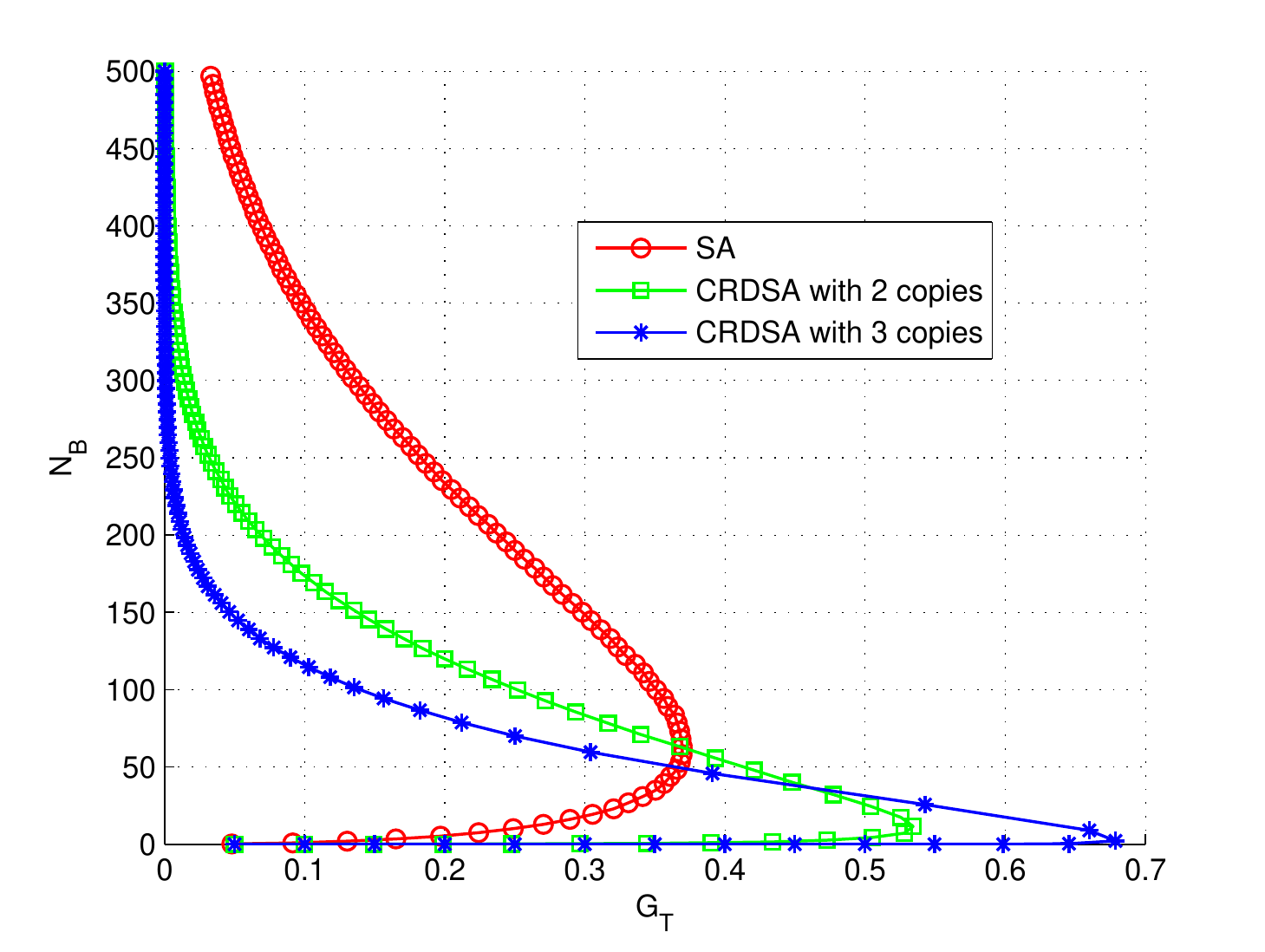}
\caption{Equilibrium contour for different packet degrees when $N_S=100$ slots}
\label{thrpcomparison}
\end{figure}

Figure~\ref{plrcomparison} shows that up to moderate channel loads, a greater packet degree is always convenient not only in terms of PLR but also in terms of throughput and packet delay. But considering stability to be one of our constraints, we can see \added{from Figure~\ref{thrpcomparison}} that for $p_r=1$ SA offers the most reliable performance since after the throughput peak any CRDSA curve rapidly degrades with $N_B$ while SA degrades in a slow-paced manner. As a result, for moderate channel load lines \replaced{with}{having} $G_T\leq0.4$ for $N_B=0$, SA is able to support up to approximately $450$ users ensuring stability, while for CRDSA with 2 replicas the admissible population is approximately $50\%$ less and for CRDSA with 3 replicas the admissible $M$ is less than 150. 

However, as previously said, the operational point for CRDSA always occurs in a point with higher throughput, smaller backlogged size and smaller packet delay. Moreover in Section~\ref{FET} it has been shown that depending on the channel load line the actual $FET$ could be so big that instability could be accepted. For example, for $M=450$ and $p_0=0.089$ the computed FET is \replaced{practically}{so big that after a long simulation (up to $f=10^{10}$) the cumulative distribution function is still equal to} zero\added{, in the sense that even though we know that a certain probability exists, it is so small that the computation resulted to be zero due to its littleness}\deleted{(thus we can assume the event of exit from the stability region really rare)}. 

Considering $M=250$ and $p_0=0.2$ SA presents a globally stable behaviour around its throughput peak ($G_{OUT}=0.36$) for $N_B^G=65$ while CRDSA with 3 replicas presents a locally stable equilibrium point with throughput equal to 0.5 , $N_B^S$ close to 0 and unstable equilibrium point for $N_B^U\approx43$. 

Using formulas in Section~\ref{FET} the obtained average FET is approximately $10^4$ frames. In this case, if we are willing to maintain the advantage\added{s} of CRDSA with 3 replicas\added{ (e.g. its lower PLR)} some countermeasures\added{ to ensure stability} are required. In the next Section we will show simple yet effective control policies able to counteract the event of drift to saturation. After that, the convenience of using such a dynamic policy in the case of finite\added{ and infinite} users' population will be discussed.

\section{Control Limit Policies}
\replaced{If we are in the case in which the considered channel is not stable, some kind of solution to counteract the possibility of drift to saturation is required. Two}{Previous sections highlighted that statistical variations can yield to instability even in case of initially well-performing channels. To render a channel of this type stable 2} straightforward solutions \replaced{could be}{can be used}: \replaced{use}{the first one is to use} a smaller value for the retransmission probability giving then rise to a larger backlogged population for the same throughput value; \replaced{allowing}{the second is to allow} a smaller user population size $M$ thus resulting in a waste of capacity. 

A third solution is the use of control limit policies to control unstable channels by applying the countermeasures above in a dynamic manner. \deleted{These dynamic control policies can be applied when it is not possible or it is not convenient to have a channel that is nominally stable.}In this section we analyze control procedures of the limit type \cite{dyn_stab} able to ensure stability in the case of stationary parameters. However the same reasoning can also be applied in the case of dynamic ones (such as non-stationary $M$) with the necessary modifications. 

Two simple yet effective dynamic control procedures are considered: the input control procedure (ICP) and the retransmission control procedure (RCP). These two control procedures are based upon a subclass of policies known as control limit policies, in which the space of the policies is generally composed of two actions plus a critical state that determines the switch between them, known as \textit{control limit}. In this case the \textit{control limit} is a critical threshold for a certain number of backlogged users $\hat{N}_B$.

\subsection{Input Control Procedure} 
This control procedure deals with new packets to transmit. In particular, two possible actions are possible: accept (action $a$) or deny (action $d$) and the switch between them is determined by the threshold $\hat{N}_B$ as previously mentioned.

\subsection{Retransmission Control Procedure}
As the name says, the retransmission control procedure deals with packets to retransmit and in particular with their retransmission probability. In particular two different retransmission probabilities $p_r$ and $p_c$ are defined that represent respectively the action taken in normal retransmission state (action $r$) and in critical state (action $c$). From the definition above it is straightforward that it must be $p_r>p_c$. The switch between these two modes is determined once again by the threshold $\hat{N}_B$.\\

\begin{figure}[tbh!]
\subfigure [ICP] {\label{ICP} \includegraphics [ scale = 0.24 ]{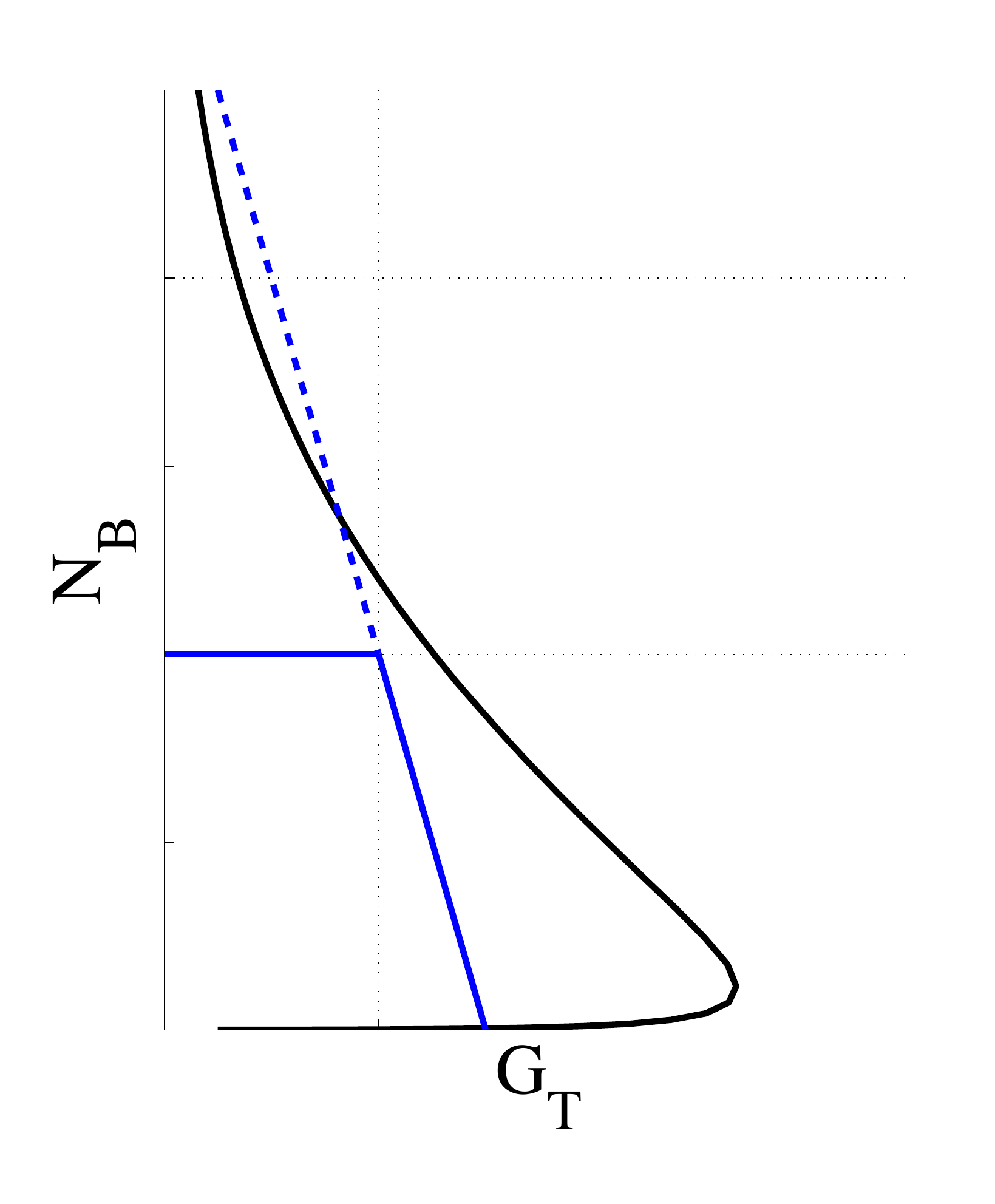}
\label{ICP}
} \qquad
\subfigure [RCP] {\label{RCP} \includegraphics [ scale = 0.24 ]{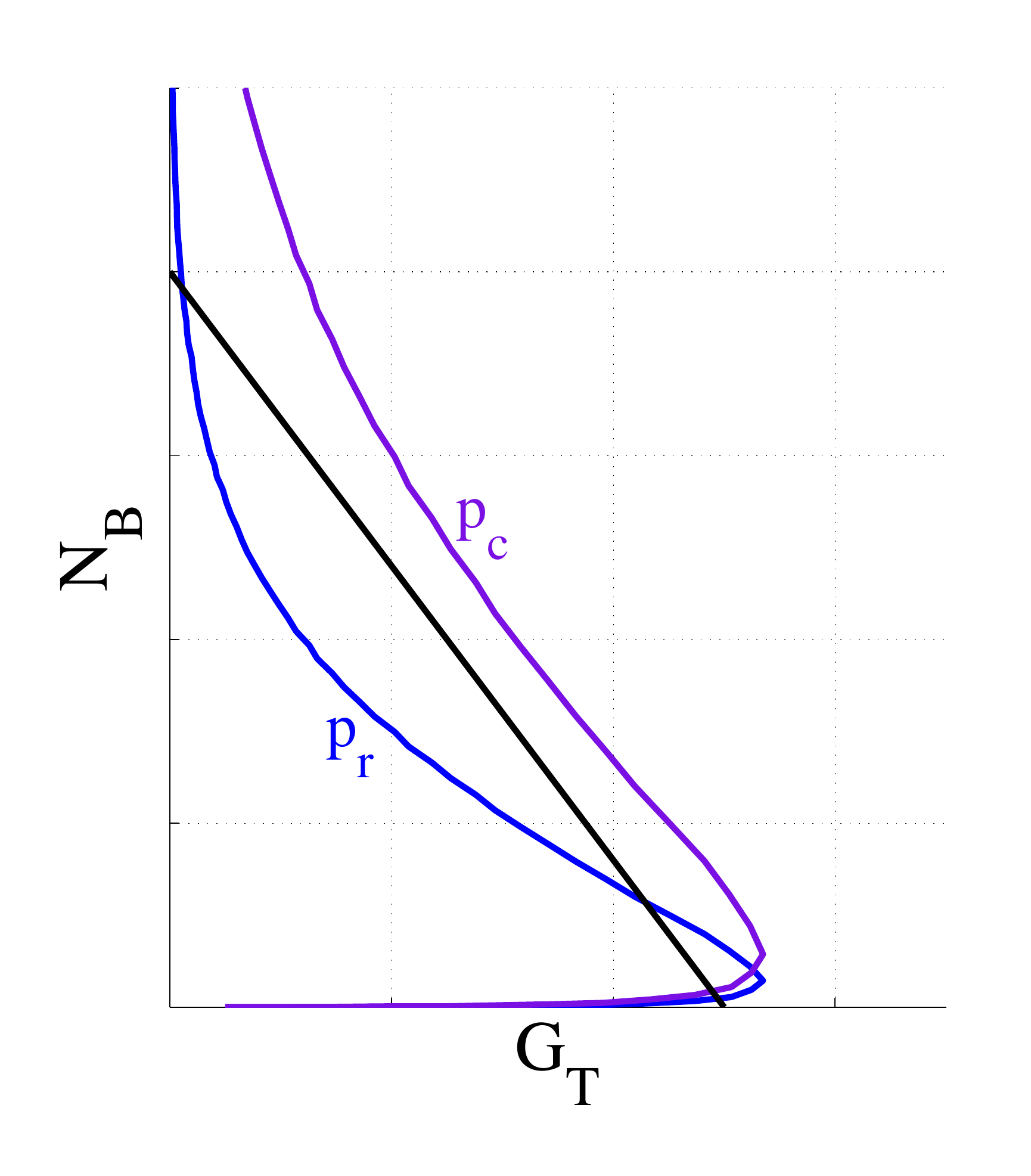}
\label{RCP}
} \qquad
\caption{Control limit policy examples.}
\label{ICPRCP}
\end{figure}

Figure~\ref{ICPRCP} graphically represents these 2 cases. As we can see both policies accomplish the same task of ensuring channel stability. However, while ICP controls the access of thinking users, RCP controls the access of backlogged users.

\section{On the use of dynamic control policies for finite population}

\begin{figure}[tbh!]
\centering
\includegraphics [width=9 cm] {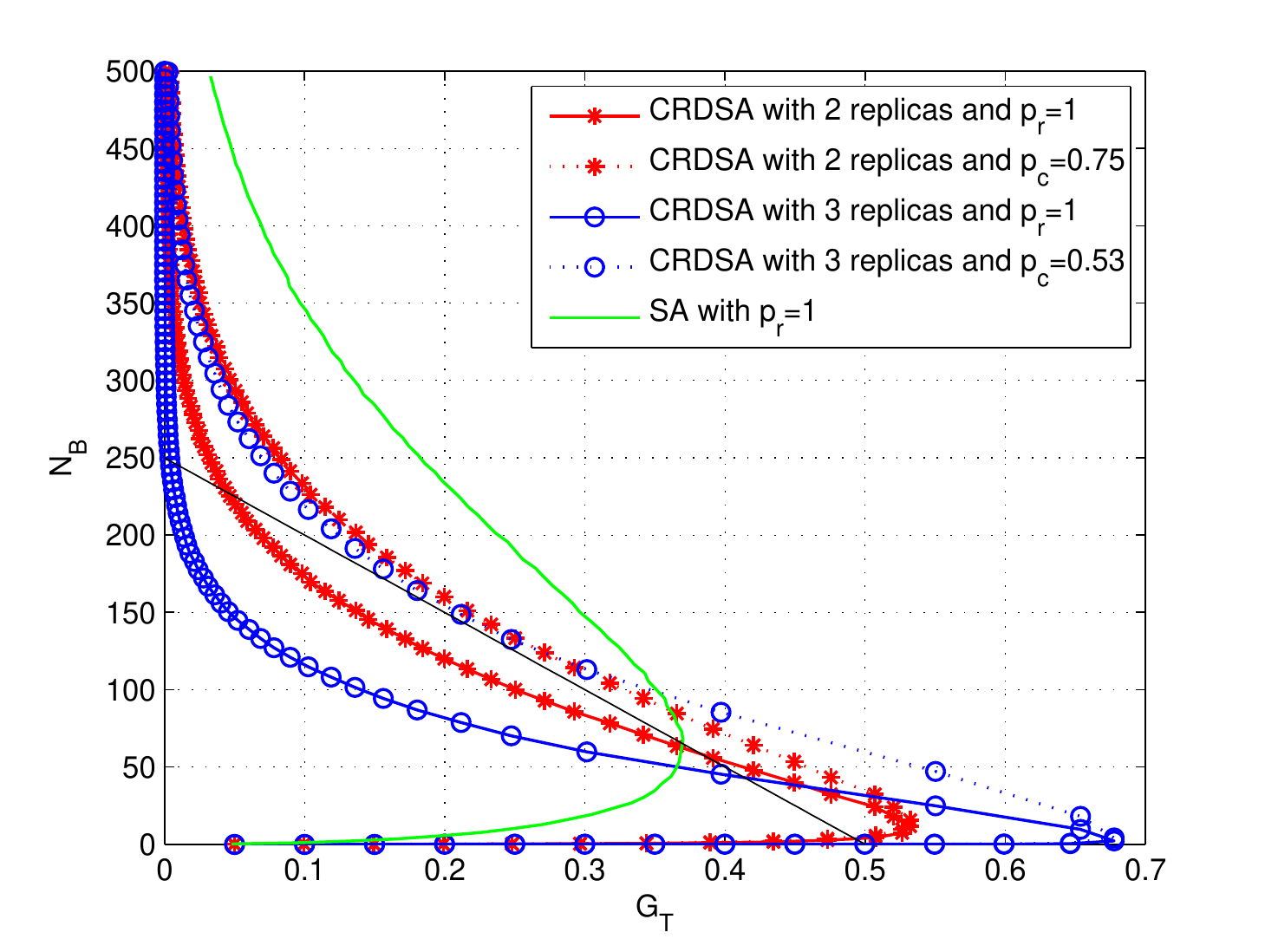}
\caption{Equilibrium contour for SA and CRDSA with 2 and 3 replicas in normal and critical RCP state}
\label{finMeq}
\end{figure}

In this section we discuss the result of the application of dynamic control policies for CRDSA with 2 and 3 replicas compared to SA and to the case of static design\deleted{ where} \added{(}$p_r=p_c$\added{)} \added{when a finite population} is assumed. Consider the case previously analyzed in Section~\ref{pktdeg} and reported in Figure~\ref{finMeq} with $M=250$, $p_0=0.2$ and $p_r=1$. Moreover we assume $p_c=0.75$ for CRDSA with 2 copies and $p_c=0.53$ for CRDSA with 3 copies, that are the biggest values for $p_c$ that ensure stability in the two respective cases.

\begin{figure}[tbh!]
\centering
\includegraphics [width=9 cm] {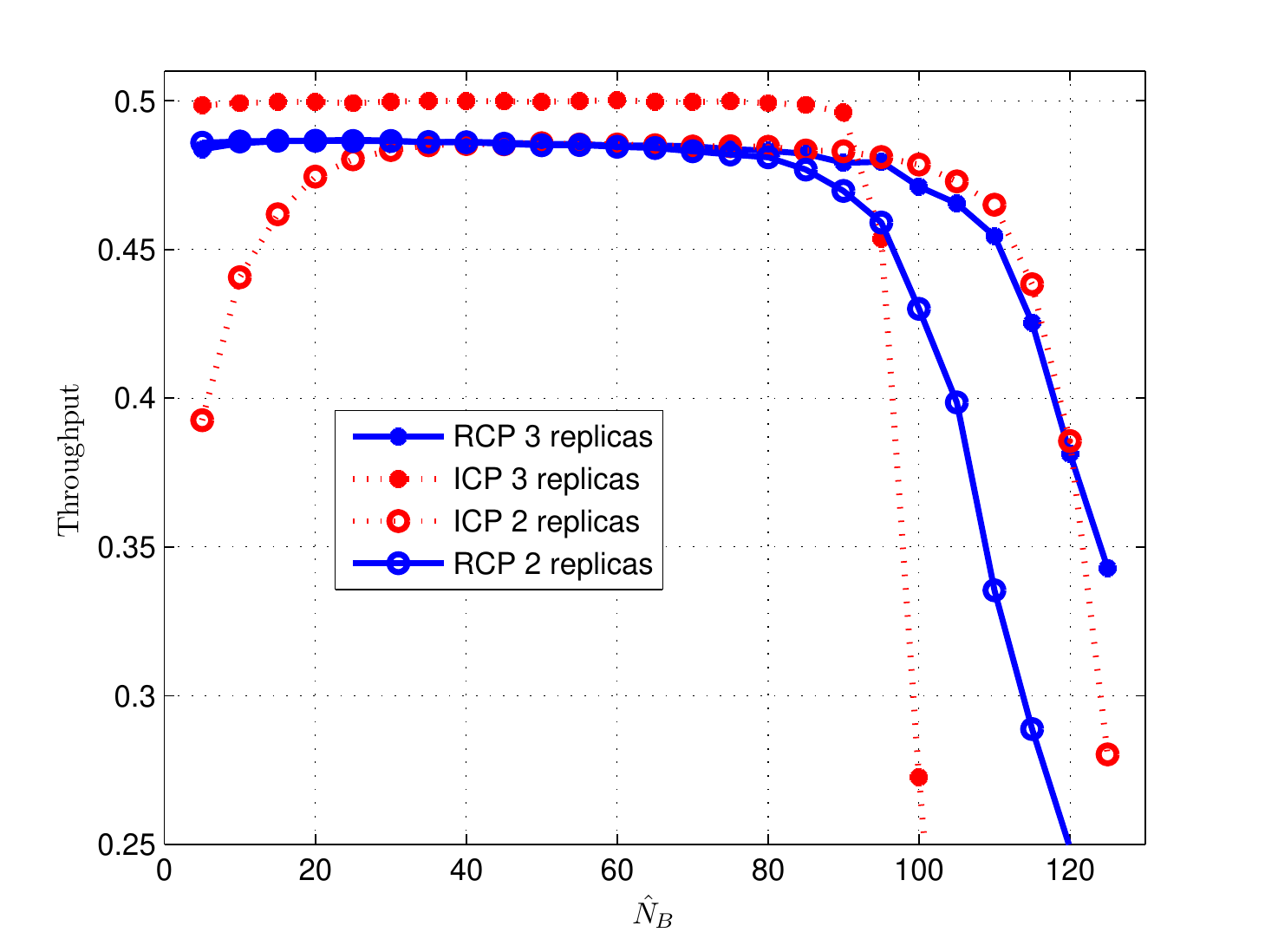}
\caption{Throughput over $\hat{N}_B$ for dynamic retransmission policies}
\label{thrpM250}
\end{figure}

\begin{figure}[tbh!]
\centering
\includegraphics [width=9 cm] {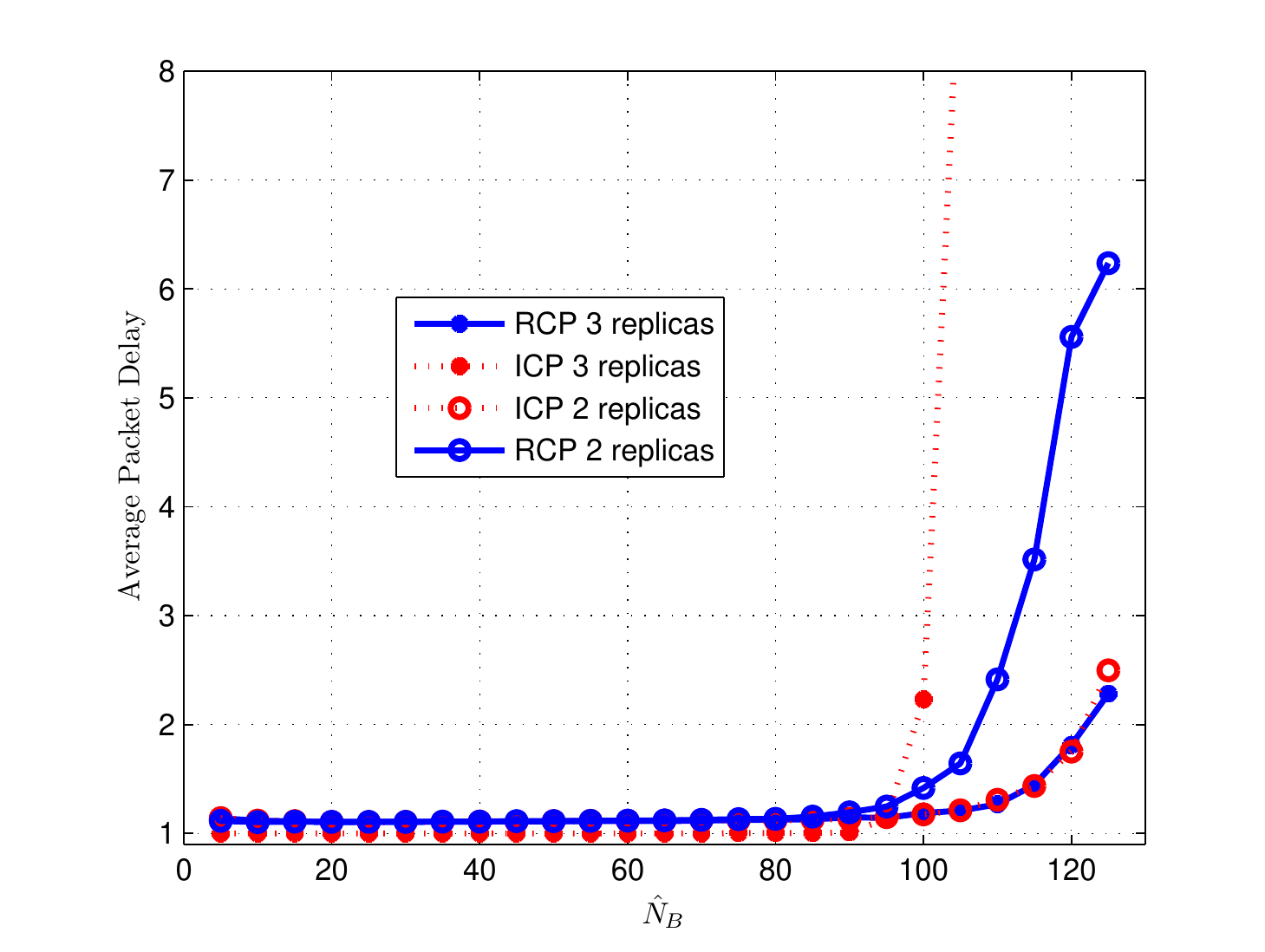}
\caption{Packet delay over $\hat{N}_B$ for dynamic retransmission policies}
\label{delM250}
\end{figure}

Figures~\ref{thrpM250} and~\ref{delM250} show the results in terms of throughput and average packet delay when varying the threshold of the control limit policy $\hat{N}_B$. As we can see the results among the various cases are similar in terms of maximum throughput and minimum delay achieved, except for ICP with 3 replicas that shows slightly higher throughput and lower delay than the other cases. 

For SA instead, the resulting channel for $M=250$ and $p_0=0.2$ does not need any dynamic control policy since it is globally stable with throughput equal to $0.369$ and delay equal to $1.76$ frames. It is now clear that despite instability that needs to be controlled, the time spent in critical state is so small that results shown for CRDSA with 2 and 3 replicas are better than SA. Therefore we can affirm that CRDSA always gets better performance than SA despite instability.  

It can also be noticed that in any case considered here, there is a certain range of flatness around $N_B^U$. This range changes depending on the policy applied, the number of replicas, the time spent in critical state and the performance degradation between normal and critical state. The width of this flat region is really important \replaced{since in}{for the following reason: imagine a real application}\added{real satellite communication} scenario\added{s}\deleted{ with} large \added{propagation} delay \replaced{has to be considered}{(as it is the case in satellite communications)}. 

\subsection{\added{Control policies in case of propagation delay}}

Differently from the case of immediate feedback assumed so far throughout the paper, when the $\hat{N}_B$ threshold is crossed the control limit policy needs a certain propagation time from the NCC to the satellite terminals before it is applied. As a result the effective threshold is not $\hat{N}_B$ but some value $\hat{N}_B+\delta$, where $\delta$ depends on the drift of the number of backlogged users for the particular case and on the propagation delay. Therefore the wider is the flat region, the less the performance optimality of the controlled channel will be influenced by the propagation delay \cite{BMSB13}. 

Moreover, we can see that in the case of 2 replicas a degradation of the throughput is also found for $\hat{N}_B<N_B^U$. In general terms, when a control limit policy is applied it does not make sense to choose $\hat{N}_B<N_B^U$, therefore this result does not seem to be of interest for our analysis. However, when $N_B<\hat{N}_B$ the control limit policy switches back to normal state, but as previously mentioned users do not immediately receive the acknowledgment for this new change due to the propagation delay so that the critical policy might actually be left for some value $\hat{N}_B-\delta$ resulting in the possibility of suboptimal performance (see ICP with 2 copies per packet).

\subsection{\added{On the use of dynamic control policies}}

Let us now discuss the results when instead of dynamic control limit policy, the channel is designed by statically decreasing the retransmission probability so that the channel is always stable. Differently from SA, in this case decrementing $p_r$ has really small influence on the overall performance especially when the first point of intersection between the equilibrium contour and the channel load line takes place before the throughput peak and sufficiently far from it. As a matter of fact with $p_r=0.53$ for CRDSA with 2 replicas and with $p_r=0.75$ for CRDSA with 3 replicas the following results are obtained:
\begin{itemize}
\item CRDSA with 2 replicas: $G_{OUT}=0.49$; $Av[D_{pkt}]=1.01$
\item CRDSA with 3 replicas: $G_{OUT}=0.5$; $Av[D_{pkt}]=1$ 
\end{itemize}

In conclusion we can say that with stationary values of $M$ and $p_0$, in most of the cases the use of dynamic policies for CRDSA does not appear to be a convenient solution compared to a static design\added{ if a choice can be made}. This can be affirmed also in light of the more realistic case of non-immediate feedback in which a certain performance degradation is present. The only case in which dynamic policies could be really necessary with stationary $M$ and $p_0$ is when $M$ is so big that the required $p_r$ rendering the channel stable really degrades the performance thus requiring the use of dynamic policies.
Nevertheless the analysis above is useful for future work on the case of non-stationary $p_0$ and $M$ that does not allow to a-priori set the retransmission probability in order to be sure that the channel is always stable. Moreover the analysis presented for the case of finite $M$ is introductory to understand the application of dynamic policies to the case of infinite population.

\section{On the use of dynamic control policies for infinite population}

Finally we discuss the use of dynamic control policies in order to provide stability in the case of infinite population. As a matter of fact this case is closer to the real DVB-over-satellite application scenario in which a big number of users is expected to generate new traffic independently of whether previous transmissions are still pending or not, so that actual channel traffic can be modeled as a Poisson process.

In the case of infinite population, ICP is the only policy that can ensure stability since decreasing $p_r$ could not be sufficient to recover stability. In fact, RCP might be used as a mean to recover the channel communication from instability by setting $p_c$ so that the operational point falls again in the region of stability. However such a procedure does not always ensure stability from a theoretical point of view since due to the satellite propagation delay, the switch from and to critical state need some time to propagate meanwhile the number of backlogged users will continue to rapidly and indefinitely increase so that at the end the chosen value for $p_c$ could not be anymore sufficient. 

\begin{figure}[t!]
\centering
\includegraphics [width=9 cm] {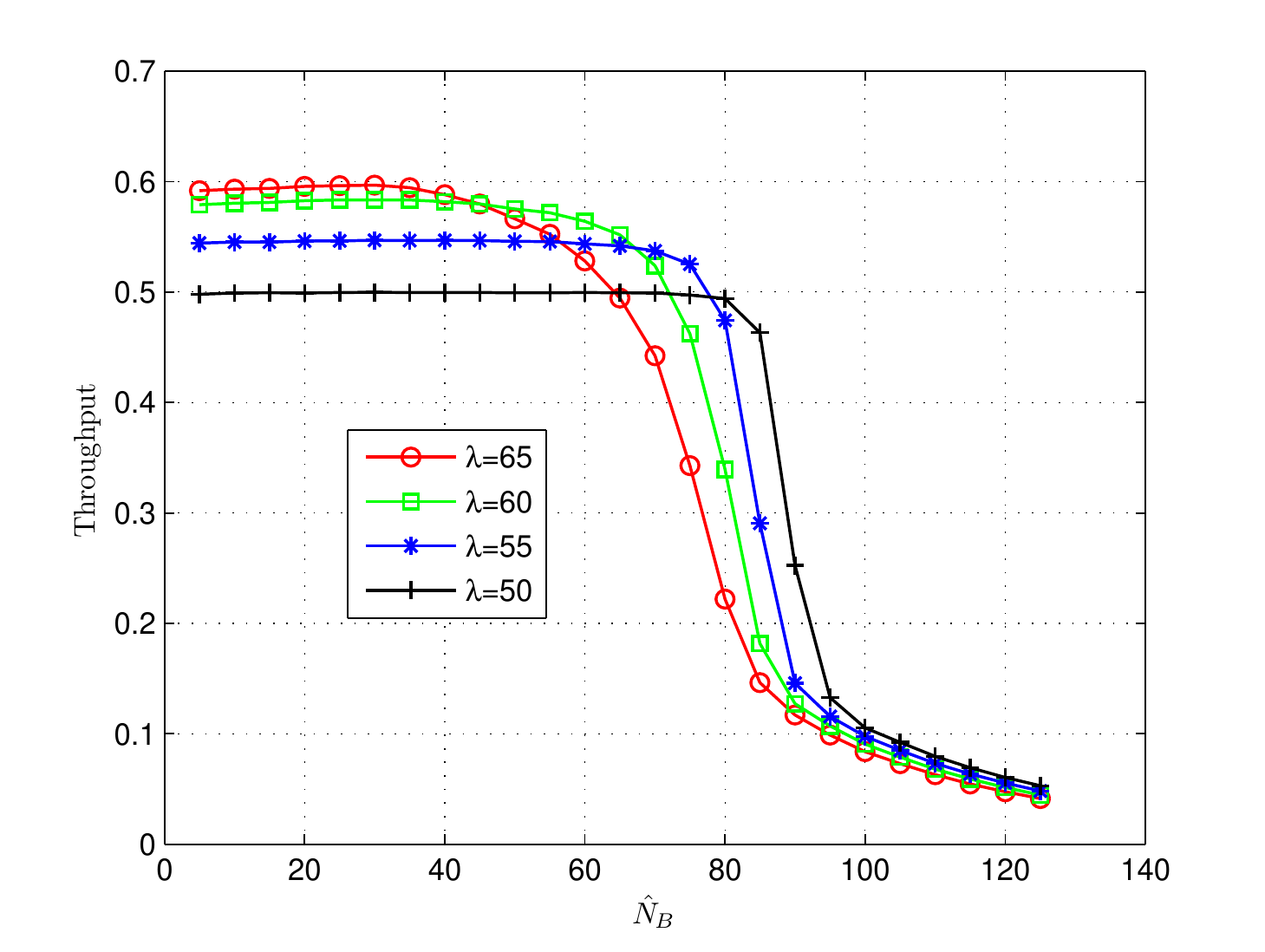}
\caption{Throughput over $\hat{N}_B$ for ICP in case of infinite population}
\label{infICPthrpTOT}
\end{figure}

\begin{figure}[t!]
\centering
\includegraphics [width=9 cm] {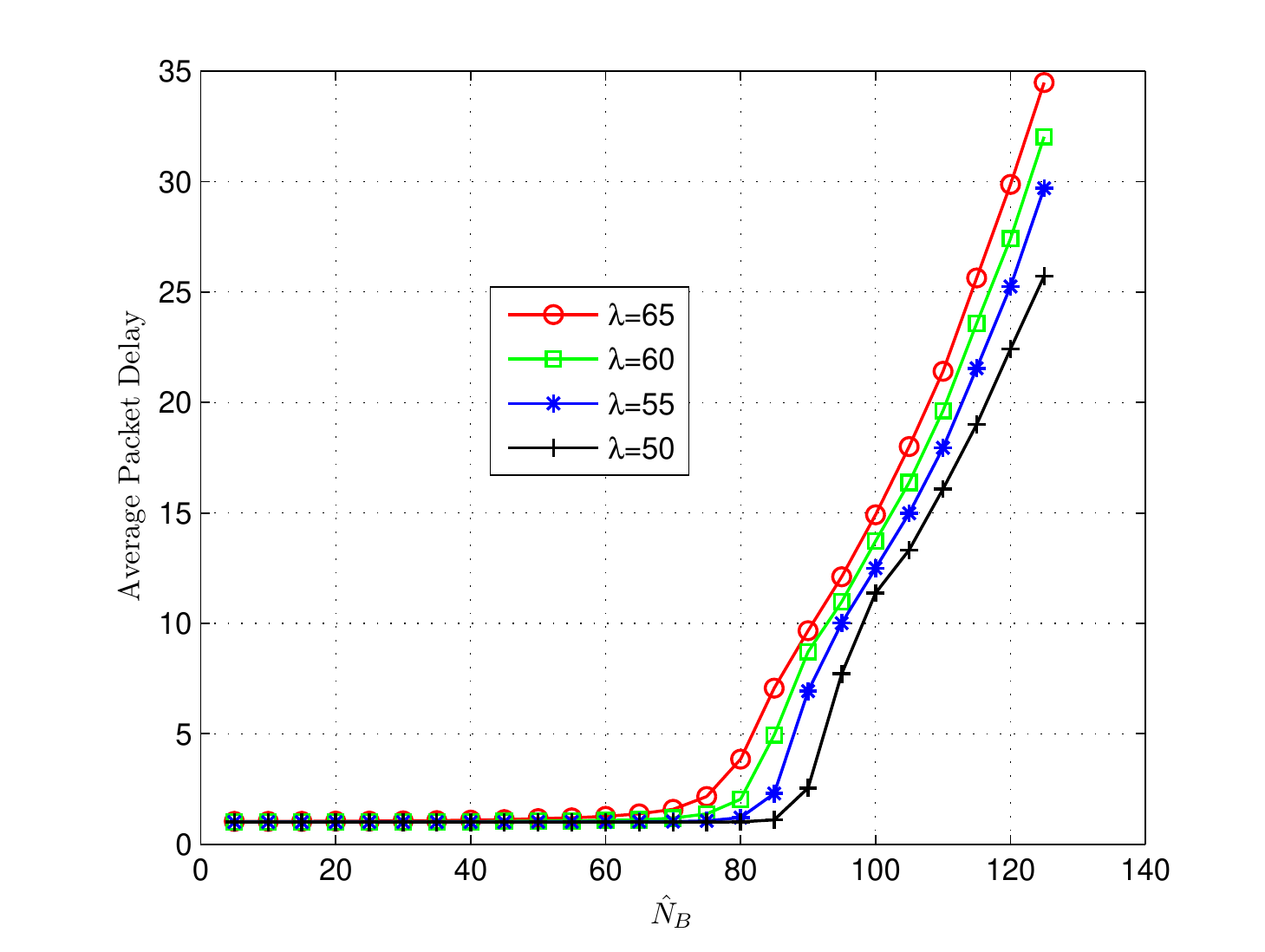}
\caption{Packet delay over $\hat{N}_B$ for ICP in case of infinite population}
\label{infICPdelTOT}
\end{figure}

\begin{figure}[t!]
\centering
\includegraphics [width=9 cm] {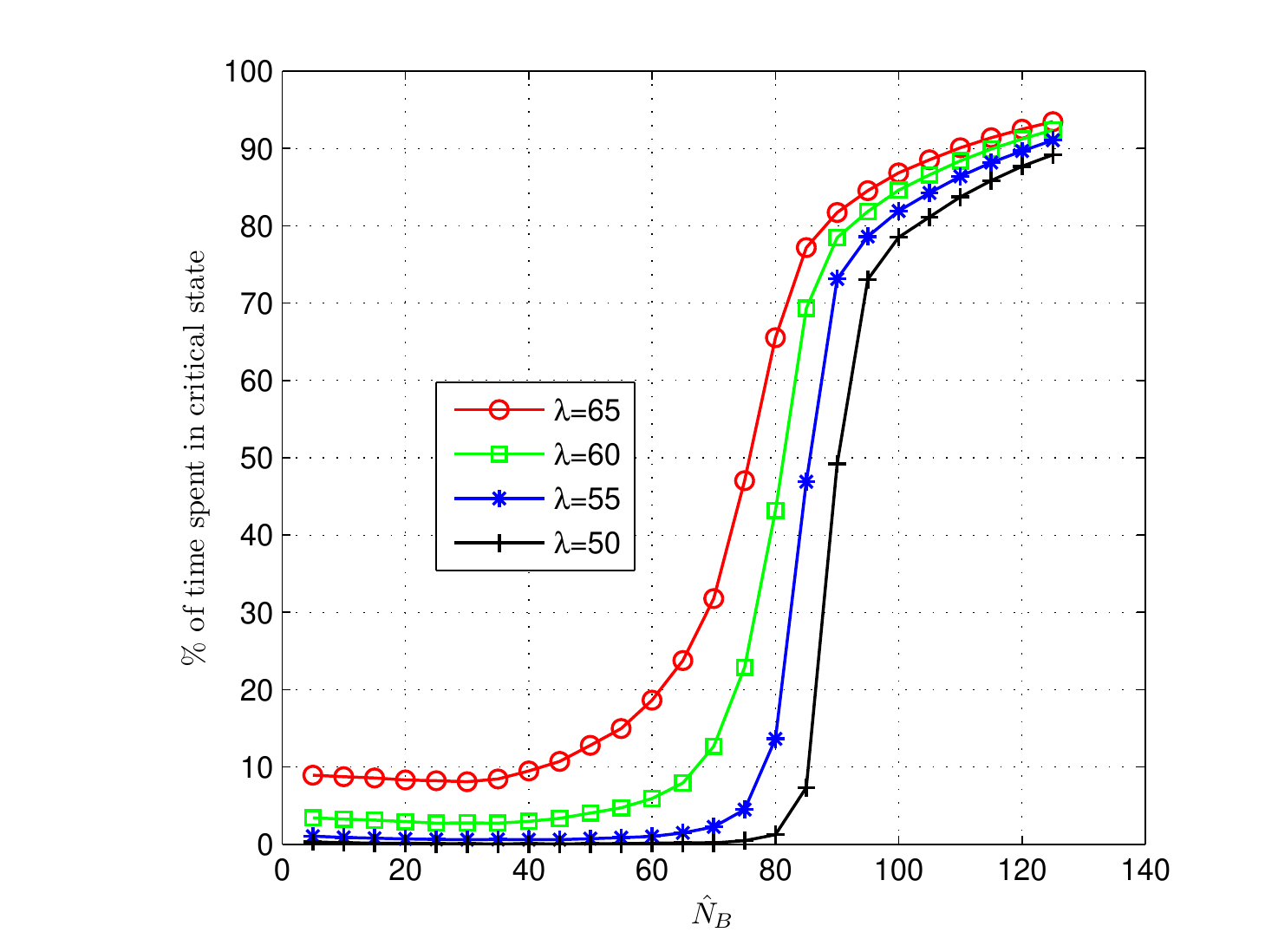}
\caption{Percentage of time spent in critical state over $\hat{N}_B$ for ICP in case of infinite population}
\label{infICPctrlTOT}
\end{figure}

\begin{table}[th!]
\centering
\begin{tabular}{| c | c | c | c |}
    \hline
    $$ & $$ & $$ & $$\\
    $\lambda=65$ & $\lambda=60$ & $\lambda=55$ & $\lambda=50$\\
    $$ & $$ & $$ & $$\\
    $N_B^U=11$ & $N_B^U=18$ & $N_B^U=25$ & $N_B^U=32$\\
    $$ & $$ & $$ & $$\\
    \hline
  \end{tabular}
\label{tab2}
\caption{Points of unstable equilibrium for CRDSA with 3 copies depending on $\lambda$}
\end{table}

Figure \ref{infICPthrpTOT} and \ref{infICPdelTOT} show throughput results depending on the number of new packet arrivals $\lambda$ at each frame. As we can see the bigger the $\lambda$ value, the higher the throughput but also the narrower the flat region. The reason for such a result is similar to what has been explained in the previous sections for the finite population case. In particular we can see from Table II that the closer the operational point is to the throughput peak, the smaller is the value of $N_B^U$. 

Figure \ref{infICPdelTOT} shows that obtained results concerning delay reflect throughput results since the wider is the flat region for the throughput, the wider is the flat region with delay equal to $1$. Finally Figure \ref{infICPctrlTOT} shows that for bigger values of $\lambda$ users spend some time in critical state also in the flat throughput region giving place on one hand to the possibility of suboptimality due to the propagation delay and on the other hand increasing the overall PLR by reason of rejected packets. 

%Finally, when the $r$ policy of ICP is applied the actual $PLR$ will increase because non-transmitted packets are surely lost. For this reason, in Figure \ref{infICPplrTOT} the difference between the PLR due to transmitted packets $PLR_tx$ and the overall one $PLR_tot$ are plot. As we can see, the result of the control limit policy application is that for high values of $\lambda$ a more frequent use of the $r$ action is required and this increases the overall PLR.
%
%\begin{figure}[t!]
%\centering
%\includegraphics [width=9 cm] {figs/infICPplrTOT}
%\caption{Packet Loss Ratio over $\hat{N}_B$ for ICP in case of infinite population}
%\label{infICPplrTOT}
%\end{figure}  

\section{Conclusions}
In this paper a model for \replaced{design of a congestion-free system}{computation of the channel stability when} using Contention Resolution Diversity Slotted Aloha as Random Access mode in DVB-RCS2 has been outlined. This model allows to study and predict stability and to calculate the expected throughput as well as the expected delay at the channel operating point. Moreover, this kind of representation allows a simple yet effective view of how the communication behaves when changing one or more of the key parameters thus allowing a design that can easily consider the tradeoff among multiple constraints. 

A simplified framework for the calculation of the First Exit Time has also been introduced and its importance for the evaluation of unstable channels has been highlighted. All the presented models and tools can be extended to consider a different frame size, power unbalance, the introduction of First Error Correction or any other modification. \deleted{However, concerning the frame size it is known from the literature that it does not bring any considerable advantage for sizes greater than 100-200 slots while the packet delay would be considerably increased.} The central part of the paper discusses design settings and the convenience of using CRDSA over SA.

In the last part, dynamic retransmission policies of the control limit type have been introduced and their effectiveness and convenience have been discussed against static design. In particular, concerning the case of finite user population, found results have shown that differently from SA in which dynamic retransmission policies are generally preferred, in the case of CRDSA a static design that ensures stability (i.e. by decreasing the retransmission probability for backlogged packets) is generally more convenient. Finally, dynamic retransmission policies have been applied to the case of infinite user population. Found results have shown that control limit policies are able to stabilize the channel also in the case of infinite user population while ensuring a performance close to the best achievable even in presence of large propagation delay as it is the case in a geostationary satellite system used for DVB-RCS communications. 

%\appendix
%Calling $G_{IN}^{tx}$ and $PLR^{tx}$ respectively the total load and the resulting $PLR$ due to transmitted packet and calling %$G_{IN}^{no-tx}$ the part of load due to new packets that has not been transmitted because of the critical state, the total $PLR$ that %takes into account also discarded packets ($PLR^{tot}$) can be written as 
%
%\begin{equation}
%PLR^{tot}=\frac{PLR^{tx}\cdot G_{IN}^{tx} + G_{IN}^{no-tx}}{G_{IN}^{tx} + G_{IN}^{no-tx}}
%\end{equation}
%
%Therefore, while the throughput for transmitted packets is
%
%\begin{equation}
%G_{OUT}^{tx}=G_{IN}^{tx}\cdot (1-PLR_{tx})
%\end{equation}
%
%if we want to consider the overall throughput
%
%\begin{equation}
%G_{OUT}^{tot}=(G_{tx}+G_{no-tx})\cdot (1-\frac{PLR_{tx}\cdot G_{tx} + G_{no-tx}}{G_{tx} + G_{no-tx}})
%\end{equation}
%
%that after some basic mathematical operations reduces to
%
%\begin{equation}
%G_{OUT}^{tot}=G_{tx}\cdot (1-PLR_{tx})=G_{OUT}^{tx}
%\end{equation}

\begin{IEEEbiography}[{\includegraphics[width=1in,height=1.25in,clip,keepaspectratio]{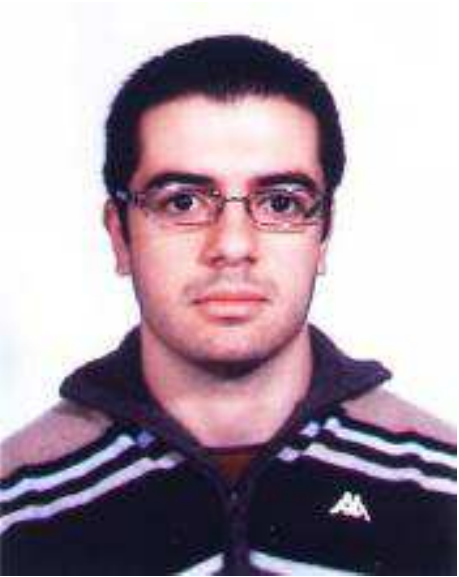}}]{Alessio Meloni}
received his M.Sc. degree in electronic engineering from the University of Cagliari (Italy) in December 2010, after spending one year at the Czech Technical University (Prague) during which he developed his thesis. Since March 2011, he has been PhD student at the Multimedia and Communications Lab of the Department of Electrical and Electronic Engineering (DIEE) - University of Cagliari. From April 2011 to October 2011 he has been guest PhD student at the German Aerospace Center, Oberpfaffenhofen, Germany. His research interests include wireless communications, advanced random access techniques and satellite links. He is a member of IEEE, IEE ComSoc, CNIT and GOLD member of BTS.
\end{IEEEbiography}

\begin{IEEEbiography}[{\includegraphics[width=1in,height=1.25in,clip,keepaspectratio]{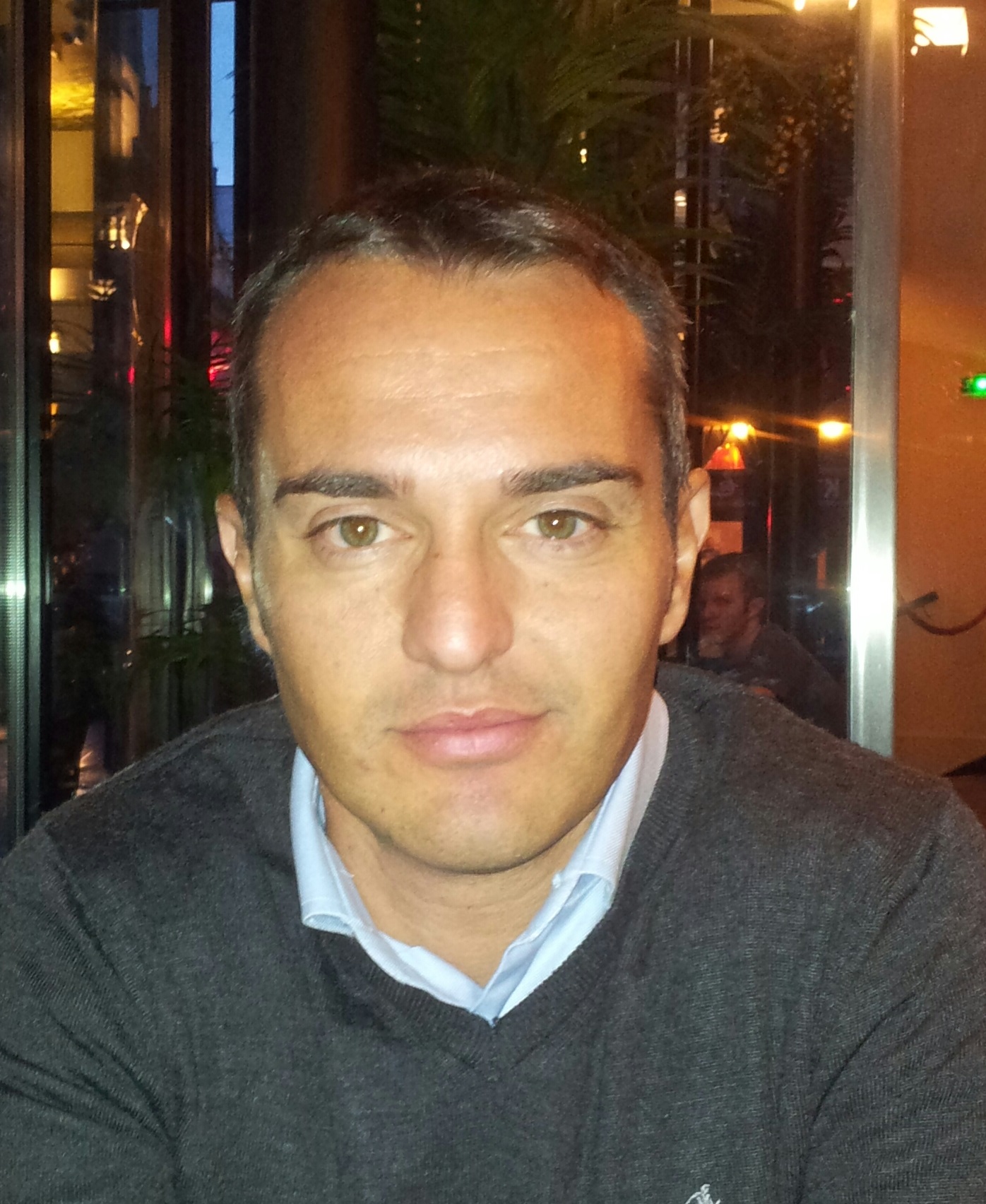}}]{Maurizio Murroni}
 (M'01) graduated (M.Sc.) with honors (Summa cum Laude) in Electronic Engineering in 1998 at the University of Cagliari and in the same year he received an award for his thesis from Telecom Italia, inc. He has been an Erasmus visiting student at CVSSP Group (Prof. Maria Petrou), School of Electronic Engineering, Information Technology and Mathematics, University of Surrey, Guildford, U.K. in 1998 and a visiting Ph.D. student at the Image Processing Group (Prof. Yao Wang), Polytechnic University, Brooklyn, NY, USA, in 2000. In 2001 he received his PhD degree in Electronic Engineering and Computers, from the University of Cagliari. In 2002 he became Assistant Professor of Communication at the Department of Electrical and Electronic Engineering (DIEE) of the University of Cagliari. Since the 1998, he contributes to the research and teaching activities of the Multimedia Communication Lab (MCLab) at DIEE. In 2006 he has been visiting professor at the Dept. of Electronics and Computers at the Transilvania University of Brasov in Romania and in 2011 at the Dept. Electronics and Telecommunications, Bilbao Faculty of Engineering, University of the Basque Country (UPV/EHU) in Spain. Since October 2010 he is coordinator of the research unit of the Italian University Consortium for Telecommunications (CNIT) at the University of Cagliari. His current research focuses on Cognitive Radio system, signal processing for radio communications, multimedia data transmission and processing. Dr. Murroni is a member of IEEE, IEEE Com Soc, IEEE BTS, IEEE DySPAN-SC and 1900.6 WG.
\end{IEEEbiography}

\end{document}